\documentclass[11pt]{article}
\pdfoutput=1
\usepackage{amsmath,amssymb,epsfig,amsfonts}
\usepackage{graphicx,subfigure}
\usepackage[usenames, dvipsnames]{color}
\usepackage[pagebackref]{hyperref}
\usepackage{cite}
\usepackage{verbatim} 
\usepackage{float}
\restylefloat{table}
\usepackage[all]{xy}
\usepackage{pdflscape}
\usepackage{tikz}
\usepackage{array}
\usepackage{youngtab}
\usepackage{multirow}
\usepackage{color}
\usepackage{ulem}

\usepackage{dsfont}

\makeatletter

\renewcommand*{\backref}[1]{}
\renewcommand*{\backrefalt}[4]{({%
    \ifcase #1 Not cited.%
          \or Page~#2.%
          \else Pages #2.%
    \fi%
    })}

\newcommand{\eqnum}{\leavevmode\hfill\refstepcounter{equation}\textup{\tagform@{\theequation}}}

\newcommand{\be}{\begin{equation}}
\newcommand{\ee}{\end{equation}}
\newcommand{\ba}{\begin{aligned}}
\newcommand{\ea}{\end{aligned}}

\newcommand{\bea}{\begin{eqnarray}}
\newcommand{\eea}{\end{eqnarray}}

\def\unit{{1\kern-.65ex {\text l}}}
\def\1{{1\kern-.65ex {\text l}}}

\def\bbP{\mathbb{P}}

\def\CC{{\cal C}}

\def\CO{{\cal O}}

\def\CQ{{\cal Q}}
\def\CR{{\cal R}}

\def\bbP{{\mathbb{P}}}
\def\bbQ{{\mathbb{Q}}}

\def\bbZ{{\mathbb{Z}}}

\def\PP1{{\mathbb{P}^1}}

\def\fkg{{\mathfrak{g}}}
\def\fku{{\mathfrak{u}}}
\def\fkf{{\mathfrak{f}}}

\newcount\hour \newcount\minute
\hour=\time \divide \hour by 60
\minute=\time
\def\now{%
\ifnum \hour<13
  \ifnum \hour=0 \advance \hour by 12 \number\hour:\else \number\hour:\fi%
     \ifnum \minute<10 0\fi%
     \number\minute%
\ A.M.%
\else \advance \hour by -12 \number\hour:%
  \ifnum \minute<10 0\fi%
  \number\minute%
  \ P.M.%
\fi%
}

\makeatother
\hypersetup{
    pdftitle={The Global Gauge Group Structure in F-theory Compactifications with U(1)s},
    pdfauthor={Mirjam Cvetic and Ling Lin},
    pdfsubject={F-theory compactifications, global gauge group structure}
}

\begin{document}

\baselineskip=18pt  
\numberwithin{equation}{section}  
\allowdisplaybreaks  



\vspace*{-2cm} 
\begin{flushright}
{\tt UPR-1286-T}\\
\end{flushright}

\vspace*{0.8cm} 
\begin{center}
 {\LARGE The Global Gauge Group Structure of\\
 F-theory Compactification with U(1)s}

 \vspace*{1.8cm}
 {Mirjam Cveti\v{c}$^{1,2}$ and Ling Lin$^{1}$}\\

\bigskip
$^1$ {\it Department of Physics and Astronomy, University of Pennsylvania, \\
209 S.~33rd Street, 
 Philadephia, PA 19104-6396, USA}\\
\smallskip
$^2$ {\it Center for Applied Mathematics and Theoretical Physics, \\
University of Maribor, Maribor, Slovenia}\\
\smallskip
  {\tt {cvetic@physics.upenn.edu, lling@physics.upenn.edu}}\\
\vspace*{0.8cm}
\end{center}
\vspace*{.5cm}
%
\noindent
We show that F-theory compactifications with abelian gauge factors generally exhibit a non-trivial global gauge group structure.
The geometric origin of this structure lies with the Shioda map of the Mordell--Weil generators.
This results in constraints on the $\fku(1)$ charges of non-abelian matter consistent with observations made throughout the literature.
In particular, we find that F-theory models featuring the Standard Model algebra actually realise the precise gauge group $[SU(3) \times SU(2) \times U(1)]/\bbZ_6$.
Furthermore, we explore the relationship between the gauge group structure and geometric (un-)higgsing.
In an explicit class of models, we show that, depending on the global group structure, an $\mathfrak{su}(2) \oplus \fku(1)$ gauge theory can either unhiggs into an $SU(2) \times SU(2)$ or an $SU(3) \times SU(2)$ theory.
We also study implications of the charge constraints as a criterion for the F-theory `swampland'.

\newpage

\tableofcontents

\section{Motivation}

F-theory \cite{Vafa:1996xn, Morrison:1996na, Morrison:1996pp} provides a beautiful connection between the physics of string compactifications and the geometry of elliptically fibred Calabi--Yau manifolds.
One of the most basic relationships is the emergence of non-abelian gauge symmetries in F-theory via singular fibres in codimension one of the fibration.
These are, according to Kodaira's and Neron's classifications \cite{MR0132556, MR0184257, MR0179172}, in one-to-one correspondence to simple Lie algebras that furnish the gauge symmetries.
By now, there exist a plethora of techniques to systematically engineer non-abelian gauge symmetries in F-theory \cite{Bershadsky:1995qy, Candelas:1997eh, Bouchard:2003bu, Katz:2011qp, Lawrie:2012gg, Braun:2013nqa, Kuntzler:2014ila, Lawrie:2014uya}.
In comparison, the geometric origin of abelian gauge symmetries associated with the Mordell--Weil group of rational sections \cite{Morrison:1996na, Morrison:1996pp, Park:2011ji, Morrison:2012ei} is much less understood.
This is in part to due the fact that sections are inherently global objects that can only be fully described within a globally defined geometry.
Consequently, there are only a handful of concrete constructions of global F-theory models with abelian gauge symmetries explicitly realised \cite{Grimm:2010ez, Krause:2011xj, Morrison:2012ei, Borchmann:2013jwa, Cvetic:2013nia, Cvetic:2013uta, Borchmann:2013hta, Cvetic:2013jta, Cvetic:2013qsa, Cvetic:2015moa, Klevers:2014bqa, Cvetic:2015ioa}.

An approach to construct models with both abelian and non-abelian gauge symmetries is to first pick a global fibration known to have sections, and then using the aforementioned techniques to introduce suitable singularities in codimension one.
This approach has been used throughout the literature to construct phenomenologically appealing models (in addition to the previous references, see also \cite{Dolan:2011iu, Mayrhofer:2012zy, Braun:2013yti, Krippendorf:2014xba, Lin:2014qga, Krippendorf:2015kta}).
However, as it is so often the case in geometry, the Mordell--Weil group of sections and codimension one singularities are not completely independent of each other.
Indeed, it turns out that the existence of torsional sections\footnote{By the Mordell--Weil theorem, the Mordell--Weil group is a finitely generated abelian group, hence must be isomorphic to $\bbZ^m \times \prod_i \bbZ_{k_i}$. Sections lying in $\prod_i \bbZ_{k_i}$ are called torsional, as opposed to those in the free part $\bbZ^m$ that give rise to abelian symmetries in F-theory.} not only enforces specific codimension one singularities corresponding to a semi-simple Lie algebra $\fkg$, it also restricts the possible matter representations \cite{Aspinwall:1998xj, Mayrhofer:2014opa} (see also \cite{Klevers:2014bqa, Oehlmann:2016wsb}).
An equivalent formulation is to say that the gauge group is not $G$---the simply connected Lie group associated to $\fkg$---but rather $G/{\cal Z}$, where ${\cal Z}$ is a subgroup of the centre of $G$.
It is important to note that in this case, only representations transforming trivially under ${\cal Z}$ are allowed.
Because field theoretically, only non-local operators such as line operators are sensitive to this quotient structure \cite{Aharony:2013hda}, one often refers to $G/{\cal Z}$ as the structure of the global gauge group, in order to distinguish it from the gauge algebra that is seen by local operators.
If ${\cal Z} \neq \{1\}$, we will refer to the global group structure to be `non-trivial'.

The analysis of \cite{Mayrhofer:2014opa} produced only models that have a non-trivial global structure in the non-abelian sector of the gauge group.
However, the central subgroup ${\cal Z}$ can also overlap with a subgroup of the abelian sector.
The most prominent example of such a non-trivial gauge group structure is in fact presumed to be the Standard Model of particle physics.
Indeed, the Standard Model spectrum is invariant under a $\bbZ_6$ subgroup that lies in the centre $\bbZ_3 \times \bbZ_2 \times U(1) \subset SU(3) \times SU(2) \times U(1)$ (for a review see, e.g., \cite{McCabe:2007zz}).
Thus the global gauge group is expected to be $[SU(3) \times SU(2) \times U(1)]/\bbZ_6$.\footnote{To be precise, the quotient could be by any subgroup of $\bbZ_6$ from the field theory perspective. Line operators differentiating between the possibilities have been recently classified in \cite{Tong:2017oea}.}
Sometimes, this global structure is seen as further evidence for an $SU(5)$ GUT, since it is a direct consequence of breaking $SU(5)$ to the Standard Model (see, e.g., \cite{Baez:2009dj}).

It may therefore seem surprising, that F-theory compactifications realising the Standard Model gauge algebra without an explicit GUT structure \cite{Lin:2014qga, Klevers:2014bqa, Cvetic:2015txa, Lin:2016vus} actually reproduces exactly the same representations which are invariant under the $\bbZ_6$ centre of $SU(3) \times SU(2) \times U(1)_Y$.
Furthermore, since there is no evidence for these models to have torsional sections, one might wonder if this agreement is purely coincidental, or if there is some further hidden structure in the geometry giving rise to the non-trivial global gauge group.

In this paper, we will show that the latter is the case.
In fact, we will present an argument---very similar to that for torsional sections in \cite{Mayrhofer:2014opa}---showing that generically, F-theory compactifications with abelian gauge factors exhibit a non-trivial gauge group structure.
We will demonstrate in section \ref{sec:general_story} that the Shioda map \cite{Shioda:1989, Shioda:1990, Wazir:2001} of sections generating the Mordell--Weil group relates the $\fku(1)$ charges of matter non-trivially to their representations under the non-abelian part of the gauge algebra.
This relationship, which leads to a non-trivial centre of the universal covering of the actual gauge group, can be equivalently understood as a refined charge quantisation condition, which has been previously observed throughout the literature.
Examples hereof will be presented in section \ref{sec:examples}, including those leading to F-theory `Standard Models'.
In section \ref{sec:higgsing}, we address the issue if and how, in F-theory, such non-trivial gauge group structures can arise from the breaking, a.k.a.~higgsing, of a larger non-abelian gauge group, similar to breaking $SU(5)$ to the Standard Model.
Because of the intricate geometric description of higgsing, we will content ourselves with the discussion of a concrete class of models having $\mathfrak{su}(2) \oplus \fku(1)$ gauge algebra.
For these, we demonstrate explicitly, how different gauge group structures arise from different breaking patterns that are captured beautifully in the geometry.
An interesting implication of our findings is presented in section \ref{sec:swampland}, where we argue that the geometric properties leading to the non-trivial global gauge group structure can also be interpreted as a criterion for effective field theories to be in the F-theory `swampland'.
This swampland criterion is formulated in terms of a charge constraint on matter representations of the non-abelian gauge algebra.
In section \ref{sec:summary}, conclusions and outlook for further investigations are presented.

\section{Shioda map and the centre of gauge groups}\label{sec:general_story}

Because our main argument is based on the Shioda map, we will first present a brief review of its prominent role in F-theory, which will also help to set up the notation.
Let $\pi: Y_{n+1} \rightarrow B_n$ be a smooth, elliptically fibred Calabi--Yau space of complex dimension $n+1$, with (singular) Kodaira fibres over a codimension one locus $\{\theta =0 \} \equiv \{\theta\} \subset B_n$, and Mordell--Weil rank $m$. 
In addition to the zero section $\sigma_0$, the Mordell--Weil (MW) group has independent sections $\sigma_k$, $1 \leq k \leq m$, which generate the free part (we will call $\sigma_k$ a `free' generator of the MW-group).
In the following, we will denote the divisor classes of the (zero) sections by ($Z$) $S_k$.
Furthermore, we have the exceptional divisors $E_i = ( \bbP^1_i \rightarrow \{\theta\} )$, $1 \leq i \leq r$, which are $\PP1$-fibrations over $\{\theta\}$.
Note that by definition, the zero section does not intersect the exceptional divisors, $Z \cdot \bbP^1_i = 0$.\footnote{Put differently, one usually defines the affine node of the generic Kodaira fibre over (an irreducible component of) $\{\theta\}$ as the one that is intersected by the zero section.}
In this set-up, the Shioda--Tate--Wazir theorem \cite{Wazir:2001} implies that the N\'{e}ron--Severi (NS) group (i.e., divisors modulo algebraic equivalence) of $Y_{n+1} \equiv Y$\footnote{Strictly speaking, the Shioda--Tate--Wazir theorem is only proven for threefolds. However, it is usually assumed in the F-theory literature that it also holds for four- and fivefolds.} satisfies
\begin{align}
	\text{NS}(Y) \otimes \bbQ = \text{span}_\bbQ ( S_1, ..., S_m ) \oplus \underbrace{\text{span}_\bbQ (Z, E_1,..., E_r) \oplus (\text{NS}(B) \otimes \bbQ)}_T \, .
\end{align}
The subspace $T$ is spanned by the zero section $Z$, the exceptional divisors $E_i$, and any divisor $D_B$ pulled back from the base $B$ with $\pi$.
Finally, let us introduce the height pairing $\langle \, , \, \rangle : \text{NS}(Y) \times \text{NS}(Y) \rightarrow \text{NS}(B)$, given by the projection $\langle D_1, D_2 \rangle = \pi (D_1 \cap D_2)$ of the intersection.

For $n \geq 2$, we know \cite{Morrison:1996na, Morrison:1996pp, Klemm:1996hh} that F-theory compactified on $Y_{n+1}$ gives rise to a gauge theory in $d=10-2n$ dimensions with gauge algebra $\fku (1)^{\oplus \, r} \oplus \fkg$ and charged matter arising from singular fibres over codimension two loci of $B_n$.
The semi-simple (non-abelian) algebra $\fkg$ is determined by the singularity types over $\{\theta\}$. In particular, the exceptional divisors $E_i$, $i= 1, ..., \text{rank}(\fkg) = r$, being dual to harmonic $(1,1)$-forms $\omega_i$, give rise---via the standard expansion $C_3 = \sum_i A_i \wedge \omega_i$ of the M-theory 3-form---to gauge fields $A_i$ taking value in the Cartan subalgebra $\mathfrak{h}$ of $\fkg$.
The W-bosons, i.e., states forming the roots of $\fkg$, originate from M2-branes wrapping the $\bbP^1_i$ fibres of $E_i$.

On the other hand, the $\fku(1)$ gauge fields arise from expanding $C_3$ along the $(1,1)$-forms $\omega_{\fku(1)_k}$ which are Poincar\'{e}-dual (PD) to divisors $\varphi(\sigma_k)$ associated with the free generators $\sigma_k$ of the Mordell--Weil group.
This so-called Shioda map \cite{Shioda:1989, Shioda:1990, Wazir:2001} is a homomorphism $\varphi: \text{MW} \rightarrow \text{NS}(Y) \otimes \bbQ$ with $\text{ker}(\varphi) = \text{MW}(Y)_\text{torsion}$, that satisfies $\langle \varphi (\sigma) , D \rangle = 0$ for any $D \in T$.
These conditions can be recast in terms of intersection numbers:
\begin{align}
	\langle \varphi(\sigma) , D_B \rangle = 0\quad & \Longleftrightarrow \quad \varphi(\sigma) \cdot \fkf = 0 \, , \label{eq:no_charge_generic_fibre}\\
	\langle \varphi(\sigma) , Z \rangle = 0 \quad & \Longleftrightarrow \quad \varphi(\sigma) \cdot \CC_B = 0 \, , \label{eq:no_flux_through_base}\\
	\langle \varphi(\sigma) , E_i \rangle = 0 \quad & \Longleftrightarrow \quad \varphi(\sigma) \cdot \bbP^1_i = 0 \, . \label{eq:no_charge_W_bosons}
\end{align}
The first two conditions ensure that the intersection product of $\varphi(\sigma)$ with the generic fibre $\fkf$ and any curve $\CC_B$ of the base (lifted by the zero section) vanishes. 
Physically, this is related to the requirement that the $\fku(1)$ gauge field lifts properly from $d-1$ to $d$ dimensions in the M-/F-theory duality.
The last condition is nothing other than the statement that the gauge bosons of $\fkg$ are uncharged under the $\fku(1)$.

These conditions determine the Shioda map up to an overall scaling: Since $\varphi$ relates a section $\sigma$ to a divisor class, we expect that $\varphi(\sigma) \sim S + $(correction terms), where $S$ is the class of $\sigma$ itself. 
To satisfy the first condition \eqref{eq:no_charge_generic_fibre}, the correction terms must contain $-Z$.\footnote{We could in fact shift by any other section $S_k$ instead of $Z$ to satisfy \eqref{eq:no_charge_generic_fibre}; however, because $\varphi$ is a homomorphism, we need $\varphi(\sigma_0) \stackrel{!}{=} 0$. Therefore, the shift has to be the divisor class $Z$ of the zero section.}
The second condition \eqref{eq:no_flux_through_base} introduces a term of the form $\pi^{-1}(D_B)$, where the exact divisor $D_B \in \text{NS}(B)$ depends on the concrete model.
We will neglect the discussion of this term, since its intersection number with any fibral curve $\Gamma$ is zero and hence does not contribute to the $\fku(1)$ charge of localised matter.
Finally, the last condition \eqref{eq:no_charge_W_bosons} gives rise to a correction term of the form $\sum_i l_i \, E_i$, where the coefficients $l_i \in \bbQ$ will be discussed in more detail momentarily.
Thus the Shioda map for any sections $\sigma$ reads
\begin{align}\label{eq:shioda_image_with_factor}
	\varphi (\sigma) = \lambda \left(S - Z + \pi^{-1} (D_B) + \sum_i l_i \, E_i \right)\, ,
\end{align}
where the overall factor $\lambda$ is not fixed by \eqref{eq:no_charge_generic_fibre} -- \eqref{eq:no_charge_W_bosons}; however, because $\varphi$ is a homomorphism, i.e., $\varphi(\sigma_1 + \sigma_2) = \varphi(\sigma_1) + \varphi(\sigma_2)$, the factor has to be the same for all sections.

Accordingly, the $\fku(1)_k$ charge of matter states which arise as M2-branes wrapping fibral curves $\Gamma$---given by the intersection number $q_k(\Gamma) = \varphi(\sigma_k) \cdot \Gamma$---are only determined up to an overall scaling, which does not have a direct physical meaning.
Therefore, we often find in the literature that the scaling is chosen such that all charges are integral.
While there is in principle nothing wrong with such a rescaling, the factor can be misleading when we analyse the global gauge group structure.
As we will see, by setting $\lambda=1$, we can read off the global gauge group directly from the coefficients $l_i$.
Field theoretically, this points towards a `preferred' $\fku(1)$ charge normalisation, in which case the $\fku(1)$ charge lattice \textit{for each representation $\CR$ of $\fkg$} has lattice spacing 1.
In this formulation, we can also interpret the non-trivial global gauge group as a relative shift by a fractional number of the charge lattice for different $\fkg$-representations.
Of course, these restrictions on the $\fku(1)$ charges have been previously observed, e.g., they are quantified in the literature \cite{Braun:2013nqa, Kuntzler:2014ila, Lawrie:2014uya, Lawrie:2015hia} for $\fkg = \mathfrak{su}(5)$, and derived more generally from the consistency of large gauge transformation in the circle reduction of F-theory \cite{Grimm:2015wda}.
The novelty of this paper is the observation that these charge restrictions are explicitly tied to the global gauge group structure for any F-theory compactification with non-trivial Mordell--Weil group.

\subsection{Fractional \textit{U}(1) charges in F-theory}\label{sec:fraction_U1_charges}

In the following, we will focus the discussion on a rank one Mordell--Weil group with a single free generator $\sigma$.
The generalisation to higher Mordell--Weil rank (and also the inclusion of torsion) is straightforward and will be presented in section \ref{sec:more_MW_generators} with explicit examples.
For the purpose of these notes, let us fix the factor $\lambda$ in the Shioda map to 1:
\begin{align}\label{eq:general_shioda_image}
	\varphi(\sigma) := S - Z + \pi^{-1}(D_B) + \sum_i l_i \, E_i \, .
\end{align}
Since the following discussion revolves around the fractional coefficients $l_i$, let us recall that they arise from requiring the intersection numbers of $\varphi(\sigma)$ with the fibre $\bbP^1_i$s of the exceptional divisors $E_i$ to vanish, see \eqref{eq:no_charge_W_bosons}.
This imposes
\begin{align}\label{eq:coefficients_Cartan_generators}
	l_i = \sum_j (C^{-1})_{ij} \, \left((S - Z + \pi^{-1}(D_B)) \cdot \bbP^1_j \right) =  \sum_j (C^{-1})_{ij} \, \left((S - Z) \cdot \bbP^1_j \right)\, . 
\end{align}
Here, $C^{-1}$ denotes the inverse of $C_{ij} = - E_i \cdot \bbP^1_j$, which is the Cartan matrix of the algebra $\fkg$.\footnote{If $\fkg = \bigoplus_a \fkg_a$, where $\fkg_a$ are simple Lie algebras, then $C$ is the block diagonal matrix formed by the Cartan matrices of $\fkg_a$.}

In general, the coefficients $l_i$ are fractional numbers that in particular depend on the intersection properties between the divisor $S-Z$ and the fibres $\bbP^1_i$ of the exceptional divisors $E_i$.
However, there is always a positive integer $\kappa$ such that $\kappa\,l_i \in \bbZ$ for all $i$. 
For example, we know that the entries of the inverse Cartan matrix of $\mathfrak{su}(n_a)$ are $z/n_a$ with $z \in \bbZ$. Hence, if $\fkg = \bigoplus_a \mathfrak{su}(n_a)$, then the smallest such $\kappa$ is the least common multiple of all $n_a$.
Note that this immediately implies charge quantisation (i.e., we really have a compact abelian gauge factor):
Since $\kappa\,\varphi(\sigma)$ is a manifestly integer class, its intersection number with fibral curves is always integral.
So the $\fku(1)$ charges (measured with respect to $\varphi(\sigma)$) of \textit{all} states realised geometrically (i.e., as M2-branes on fibral curves) lie in a lattice of spacing $1/\kappa$.

In fact, the Shioda map \eqref{eq:general_shioda_image} makes an even more refined statement.
Because $S$ and $Z$ are divisor classes of sections, they are manifestly integer, i.e., their intersection product with fibral curves $\Gamma$ must be integer as well.
But then, the charge of the matter state ${\bf w}$ associated with $\Gamma$ must satisfy
\begin{align}\label{eq:integer_pairing_condition}
\begin{split}
	& q^{\bf w} = \varphi(\sigma) \cdot \Gamma = \left( S - Z + \sum_i l_i \, E_i \right) \cdot \Gamma \\
	\Longrightarrow \quad & q^{\bf w} - \sum_i l_i \, E_i \cdot \Gamma = q^{\bf w} - \sum_i l_i \, {\bf w}_i = (S - Z) \cdot \Gamma \in \bbZ \, ,
\end{split}
\end{align}
where, in the second line, we have used the standard result that the Dynkin labels of a weight ${\bf w}$ associated with a fibral curve $\Gamma$ is given by ${\bf w}_i = E_i \cdot \Gamma \in \bbZ$.
Curves $\Gamma_{\bf w,v}$ localised at the same codim 2 locus, but realising different states ${\bf w,v}$ of the same $\fkg$-representation, differ by an integer linear combination $\mu_k \, \bbP^1_k$s, since these $\bbP^1$s correspond to the simple roots of the algebra $\fkg$.\footnote{\label{footnote:same_charge_for_one_rep}
By \eqref{eq:no_charge_W_bosons}, these states must have the same $\fku(1)$ charge, hence form a single representation $(q^{\cal R}, {\cal R}_\fkg)$ of the full algebra $\fku(1) \oplus \fkg$.
} 
For these, we have
\begin{align}
\begin{split}
	\sum_i l_i \, {\bf w}_i  = & \sum_i l_i  \, E_i \cdot \Gamma_{\bf w} = \sum_i l_i \, E_i \cdot (\Gamma_{\bf v} + \sum_k \mu_k \, \bbP^1_k) = \sum_i \left( l_i \, {\bf v}_i - \sum_k \mu_k \, l_i C_{ik} \right) \\
	\stackrel{\text{\eqref{eq:coefficients_Cartan_generators}}}{=} & \sum_i l_i \, {\bf v}_i - \sum_k \underbrace{ \mu_k \, (S-Z)\cdot \bbP^1_k}_{\in \bbZ} \, .
\end{split}
\end{align}
Thus, we can associate to each $\fkg$-representation ${\cal R}_\fkg$ a $\kappa$-fractional number between 0 and 1,
\begin{align}\label{eq:L_of_rep}
	L({\cal R}_\fkg) := \sum_i l_i \, {\bf w}_i \! \mod \bbZ \, ,
\end{align}
which is independent of the choice of weight ${\bf w} \in {\cal R}_\fkg$.
For a representation $(q^{\cal R}, {\cal R}_\fkg)$ of $\fku(1) \oplus \fkg$, this allows to rewrite \eqref{eq:integer_pairing_condition} as a condition for the $\fku(1)$ charge,
\begin{align}\label{eq:integer_condition_charges_rep}
	q^{\cal R} - L({\cal R}_\fkg) \in \bbZ \, .
\end{align}
So for any matter with $\fkg$-representation $\CR_\fkg$, the possible $\fku(1)$ charges arrange in a lattice of integer spacing.
However, for different representations, the lattices do not in general align.
In fact, from what we have seen above, they can differ by multiples of $1/\kappa$.
%

The geometric origin of \eqref{eq:integer_condition_charges_rep} lies in the intersection properties of divisors and codimension one singular fibres over $\{\theta\}$.
Indeed, the non-integrality of the coefficients $l_i$ \eqref{eq:coefficients_Cartan_generators}, which leads to the non-trivial integrality condition \eqref{eq:integer_pairing_condition}, stems from the zero section $Z$ and the generating section $S$ intersecting $\bbP^1$ fibres of possibly different exceptional divisors $E_i$.
This so-called \textit{split} \cite{Braun:2013yti, Kuntzler:2014ila} of the fibre structure over $\{\theta\}$ by the section can be easily determined in concrete models, e.g., directly from the polytope in toric constructions \cite{Braun:2013nqa}.
The analysis carried out to obtain \eqref{eq:integer_condition_charges_rep} is essentially equivalent to the study of the fibre splitting patterns in the presence of sections and the allowed $\fku(1)$ charges, e.g., as in \cite{Kuntzler:2014ila,Lawrie:2015hia} for classifying all possible $\fku(1)$ charges of $\mathfrak{su}(5)$ matter.
Here, we have rephrased it in a way that allows for a more straightforward connection to the global structure of the gauge group.
An alternative way of deriving \eqref{eq:integer_pairing_condition} is to consider circle compactifications of F-theory and require consistency of the large gauge transformations along the circle \cite{Grimm:2015wda}.

Note that the above discussion, in particular the derivation of \eqref{eq:integer_condition_charges_rep} for matter localised in codimension two, is based purely on codimension one properties.
Hence, all arguments and conclusions hold for F-theory compactifications to six, four and two dimensions.

\subsection{Non-trivial central element from the Shioda map}

To see how the above observation relates the Lie algebra $\fku(1) \oplus \fkg$ to the global gauge group $G_\text{glob}$, first note that $G_\text{glob}$ has $U(1) \times \tilde{G}$ as a cover, where $\tilde{G}$ is the simply connected Lie group associated to $\fkg$.
We can now define an element of the centre ${\cal Z}(U(1) \times \tilde{G}) = U(1) \times {\cal Z}(\tilde{G})$, which has to act trivially on all geometrically realised weights.
For that, we first define the element $\Xi := q - \sum_i l_i \, E_i$ of the Cartan subalgebra $\fku(1) \oplus \mathfrak{h} \subset \fku(1) \oplus \fkg$, where $q$ is the generator of $\fku(1)$.
Its action on the representation space of an irreducible representation ${\CR} = (q^{\CR}, {\CR}_\fkg)$ of $\fku(1) \oplus \fkg$ is then simply defined through its action on the weights ${\bf w} \in \CR$.
Explicitly, denoting by ${\bf w}_i$ the Dynkin labels of ${\bf w}$ under $\fkg$, we have
\begin{align}
	\Xi ({\bf w}) :=  q^{\CR}\,{\bf w} - \left( \sum_i l_i \, {\bf w}_i \right) \times \mathds{1} \, {\bf w} \, ,
\end{align}
where $\mathds{1}$ is the identify matrix in the representation ${\CR}_\fkg$.
By exponentiating this equation, we obtain the action of a group element in $U(1) \times \tilde{G}$ on weights of $\CR$,
\begin{align}\label{eq:central_element_from_Xi}
\begin{split}
	C \, {\bf w} := & \exp \left( 2\pi  i \, \Xi \right) {\bf w} = \left[  \exp (2\pi i\, q^{\CR} ) \otimes \left( \exp( - 2\pi i \, \sum_i l_i \, {\bf w}_i ) \times \mathds{1}  \right) \right] {\bf w} \\
	\stackrel{\text{\eqref{eq:L_of_rep}}}{=} & \left[  \exp (2\pi i\, q^{\CR} ) \otimes \left( \exp( - 2\pi i \, L({\cal R}_\fkg)  ) \times \mathds{1}  \right) \right] {\bf w} \, .
\end{split}
\end{align}
Evidently, $C$---being proportional to the identify element of $\tilde{G}$---commutes with every element, i.e., $C$ is in the centre $U(1) \times {\cal Z}(\tilde{G})$.
Let us now restrict the action to representations realised in the F-theory compactification, i.e., weights $\bf w$ that arise from fibral curves $\Gamma$.
Because the tensor product\footnote{The tensor product arises, because any finite dimensional irreducible representation of a product group is a tensor product of irreducible representations of the factors.} is bilinear, the expression can be also written as
\begin{align}\label{eq:trivial_action_of_Xi}
	C \, {\bf w}  = \left[ \vphantom{\sum_i} \right. \exp [ 2\pi i \, (\underbrace{q^{\CR} - \sum_i l_i \, {\bf w}_i}_{\in \bbZ \, \text{ from \eqref{eq:integer_pairing_condition}}} ) ] \otimes \mathds{1} \left. \vphantom{\sum_i} \right]  {\bf w} = {\bf w} \, ,
\end{align}
i.e., $C$ acts trivially on weights ${\bf w}$ arising from fibral curves $\Gamma$!
But on the other hand, we also know from the previous discussion that there is a positive integer $\kappa$ such that $\kappa \, l_i \in \bbZ$ for all $i$, see paragraph after \eqref{eq:coefficients_Cartan_generators}.\footnote{It is implicitly assumed that we choose $\kappa$ to be the smallest positive integer such that $\kappa\,l_i \in \bbZ$ for all $i$.}
Going back to \eqref{eq:central_element_from_Xi}, for which we introduce the short-hand notation
\begin{align}\label{eq:central_element_from_Xi_center_form}
	C \, {\bf w} = [Q^{\CR} \otimes (\xi_{\bf w} \times \mathds{1})] \, {\bf w} \, ,
\end{align}
we see that for weights in a \textit{any} representation of $\tilde{G}$, we have
\begin{align}
\begin{split}
	& (\xi_{\bf w} \times \mathds{1})^\kappa \equiv (\exp( 2\pi i \, l_i \, {\bf w}_i) \times \mathds{1})^\kappa = \exp( 2\pi i \,\kappa\, l_i \, {\bf w}_i) \times \mathds{1} = \mathds{1} \, .
\end{split}
\end{align}
In other words, $\xi_{\bf w} \times \mathds{1}$ generates a $\bbZ_\kappa$ subgroup of the centre ${\cal Z}(\tilde{G})$.
But because we have shown that all states in geometrically realised representations must be acted on trivially by $Q^{\bf w} \otimes (\xi_{\bf w} \times \mathds{1})$, we conclude that the global gauge group structure should be
\begin{align}\label{eq:global_gauge_group_structure_general}
	G_\text{glob} = \frac{U(1) \times \tilde{G}}{ \langle C \rangle} \cong \frac{U(1) \times \tilde{G} }{\bbZ_\kappa} \, .
\end{align}
As mentioned before, the whole discussion applies to F-theory compactified to six, four and two dimensions.

Note that, strictly speaking, the second equality in \eqref{eq:global_gauge_group_structure_general} is merely a definition of the notation $[U(1) \times \tilde{G}] / \bbZ_\kappa$.
Indeed, from a purely representation theoretic point of view, we do not know that $(Q^{\bf w})^\kappa = 1$ for \textit{every} charged state (the charges could be quantised finer than $1/\kappa$).
However, we have seen above that the geometry of the F-theory model actually dictates the charges to be quantised in units of $1/\kappa$, i.e., $C^\kappa = \text{id}$.
In our discussion, both charge quantisation and the central element $C$ follow from the same observation, namely the integrality condition \eqref{eq:integer_pairing_condition}.
Hence, the notation $(U(1) \times \tilde{G})/\bbZ_\kappa$ can also be seen as encoding the $\fku(1)$ charge quanta of an F-theory compactification.

The reader might recognise the above argument, leading up to \eqref{eq:central_element_from_Xi}, from \cite{Mayrhofer:2014opa}, which related the presence of $\kappa$-torsional sections to the $\bbZ_\kappa$-center of purely non-abelian groups (i.e., no $U(1)$ factor in the cover of $G_\text{glob}$).
Indeed, there one arrives by the same logic at \eqref{eq:central_element_from_Xi_center_form} with $Q = 1$. In that case the conclusion is simply that $\xi_{\bf w} \times \mathds{1}$, which generates a subgroup of the centre ${\cal Z}(\tilde{G})$, must act trivially.

Finally, we note that even though the above discussion has been limited to a single $\fku(1)$ factor, the analysis readily extends to multiple sections $\sigma_k$ (free or torsional).
Because the Shioda map \eqref{eq:general_shioda_image} of any Mordell--Weil generator $\sigma$ (free or torsional) takes the form $S - Z +(\text{non-sectional divisors})$, one quickly realises that each Mordell--Weil generator $\sigma_k$ gives rise to an independent trivially acting central element $C_k$.
Thus, the global gauge group structure is a quotient by a product of $\bbZ_{\kappa_k}$'s.
We will come back to explicit examples hereof in section \ref{sec:more_MW_generators}.

\subsection{Preferred charge normalisation in F-theory}\label{sec:preferred_charge_normalisation}
%

Let us revisit the possible rescaling \eqref{eq:shioda_image_with_factor} of the Shioda map and the resulting normalisation of $\fku(1)$ charges in F-theory.
In field theory, the overall scaling of the $\fku(1)$ charge is unphysical, and can be chosen to our convenience.
Likewise, as mentioned before, the Shioda map is only defined up to a constant rescaling.
However, in F-theory we have a preferred normalisation provided by the integer divisor classes of sections.

Explicitly, given free generator $\sigma$, we know that its divisor class $S$ must be integer and intersecting the generic fibre once. Any rescaling of $S$ cannot preserve these properties.
Furthermore, if we rescale $\varphi(\sigma) = S - Z + l_i\,E_i$ by an integer $\kappa$, then, depending on the non-abelian gauge algebra and the fibre split structure, $\kappa\,l_i$ could be integer, which makes the $\fku(1)$ generator potentially `blind' to the central element $C$.
Indeed, if we were to repeat the analysis leading to \eqref{eq:integer_pairing_condition} with the divisor $\kappa \, \varphi(\sigma)$, then the equivalent expression becomes
\begin{align*}
	q^\text{w}_\kappa - \sum_i \underbrace{\kappa\,l_i}_{\in \bbZ} \,{\bf w}_i = \kappa \, (S - Z) \cdot \Gamma \in \bbZ \, ,
\end{align*}
which does not provide any non-trivial relation between the $\fku(1)$ charge and the weight vectors ${\bf w}$.
On the other hand, if we rescale the $\fku(1)$ charge by a fractional number $\lambda$, then it is no longer guaranteed that $\lambda \, (S - Z) \cdot \Gamma$ is always an integer.

Therefore, it is only with the normalisation $\varphi(\sigma) = S - Z +...$ of the $\fku(1)$ generator, that we can make the non-trivial relation \eqref{eq:integer_pairing_condition} manifest in any F-theory compactification.
Comparing to the field theory perspective, where any rescaling of $\fku(1)$ charges has no physical meaning, we conclude that the appropriate field theoretic data associated with this preferred normalisation are the global gauge group structure and the charge quantisation of individual $\fkg$-representations.
Equivalently, by first establishing these data, one is then free to choose any normalisation for the $\fku(1)$ charge in the field theory.

\section{The global gauge group of F-theory models}\label{sec:examples}

In this section, we will apply the above analysis to concrete models with $\fku(1)$s that have been constructed over the last few years in the literature.

\subsection[Models with \texorpdfstring{$\mathfrak{su}(5) \oplus \fku(1)$}{su(5)+u(1)} singularity]{Models with \boldmath{$\mathfrak{su}(5) \oplus \fku(1)$} singularity}\label{sec:su5+u1_examples}

Let us begin with arguably one of the most studied F-theory model, namely the so-called $U(1)$-restricted Tate model \cite{Grimm:2010ez} given by the hypersurface
\begin{align}\label{eq:restricted_Tate_model}
	y^2 + a_1\,x\,y\,z + a_3\,y\,z^3 = x^3 + a_2\,x^2\,z^2 + a_4\,x\,z^4 \, .
\end{align}
The origin of the $\fku(1)$ symmetry in the restricted Tate model can be traced to the appearance of a rational section
\begin{align}
	\sigma : [x:y:z] = [0:0:1]
\end{align}
with divisor class $S$, in addition to the standard zero section $\sigma_0 : [x:y:z] = [1:1:0]$ with class $Z$ \cite{ Krause:2011xj}.
By tuning the coefficients $a_i$ following Tate's algorithm \cite{MR0393039, Bershadsky:1996nh, Katz:2011qp}, $a_2 = a_{2,1}\, \theta , \, a_3 = a_{3,2} \,\theta^2, \, a_4 = a_{4,3}\,\theta^3$, the elliptic fibration \eqref{eq:restricted_Tate_model} develops an $\mathfrak{su}(5)$ singularity over $\{\theta\}$.
The resolution of this singularity introduces four exceptional Cartan divisors $E_i$, of which only $E_3$ is intersected once by $S-Z$.\footnote{
Recall that the zero section $Z$ intersects the affine node of $I_5$ fibre over $\{\theta\}$. The affine node is separated from the $\bbP^1_3$ node by the fibre of $E_4$. In the notation of \cite{Kuntzler:2014ila}, this is a fibre split type $(0||1)$.
}
Inserting into \eqref{eq:coefficients_Cartan_generators} then yields $l_i = \frac{1}{5}(2,4,6,3)_i$.
The corresponding central element \eqref{eq:central_element_from_Xi} generates a subgroup $\bbZ_5 \subset SU(5) \times U(1)$, which has to act trivially on representations realised geometrically.
Hence, the global gauge group of the $U(1)$-restricted Tate model with $\mathfrak{su}(5)$ singularity must be $(SU(5) \times U(1))/\bbZ_5$.
Note that for these values of $l_i$, \eqref{eq:L_of_rep} yields $L({\bf 10}) = \frac{4}{5}$ and $L({\bf 5}) = \frac{2}{5}$.
Hence, by \eqref{eq:integer_condition_charges_rep}, any ${\bf 10}$ representation must have $\fku(1)$ charge $\frac{4}{5} \mod \bbZ$, while any ${\bf 5}$ representation has charge $\frac{2}{5} \mod \bbZ$.
This is of course consistent with the spectrum, which in terms of the normalised $\fku(1)$ generator $\varphi(\sigma) = S - Z + l_i\,E_i$ reads
\begin{align*}
	{\bf 10}_{-1/5} \, , \quad {\bf 5}_{-3/5} \, , \quad {\bf 5}_{2/5} \, , \quad {\bf 1}_1 \, .
\end{align*}

Note that there are also $\mathfrak{su}(5) \oplus \fku(1)$ models with a $\bbZ_5$ centre that is embedded differently into the $U(1)$, leading to different charge assignments.
One such example can be constructed via `toric tops' \cite{Candelas:1996su, Bouchard:2003bu} in a $\text{Bl}_1 \bbP_{112}$-fibration \cite{Morrison:2012ei}.
It is labelled `top 2' in the appendix of \cite{Borchmann:2013jwa}, which is equivalent to the model `$\CQ(4,2,1,1,0,0,2)$' in \cite{Kuntzler:2014ila}.
Without going into the details of this model, we note that the sections $Z$ and $S$ intersect in neighbouring nodes of $\mathfrak{su}(5)$ fibre (i.e., fibre split type $(0|1)$ in the notation of \cite{Kuntzler:2014ila}).
The Shioda map is then $S - Z + \frac{1}{5} (1, 2, 3, 4)_i \, E_i$, which also leads to a $\bbZ_5$ centre.
However, the $\fku(1)$ charges are constrained to be $\frac{1}{5} \! \mod \bbZ$ for ${\bf 5}$-matter and $\frac{2}{5} \! \mod \bbZ$ for ${\bf 10}$-matter.
Correspondingly, the spectrum reads
\begin{align*}
	{\bf 10}_{2/5} \, , \quad {\bf 5}_{6/5} \, , \quad {\bf 5}_{-4/5} \, , \quad {\bf 5}_{1/5} \, , \quad {\bf 1}_1 \, , \quad {\bf 1}_2 \, .
\end{align*}

Of course there are also models without a non-trivial global gauge group structure.
An example is the model labelled `top 4' in the appendix of \cite{Borchmann:2013jwa}, or `$\CQ(3,2,2,2,0,0,1)$' in \cite{Kuntzler:2014ila}.
Here, both sections $Z$ and $S$ intersect the affine node of $\mathfrak{su}(5)$ (i.e., fibre split type (01)). So the Shioda map is $\varphi(\sigma) = S-Z$, without any shifts by Cartan divisors.
The central element \eqref{eq:central_element_from_Xi} then just imposes that all charges must be integral.
Thus, the global gauge group is $SU(5) \times U(1)$, which of course is consistent with the spectrum
\begin{align*}
	{\bf 10}_0 \, , \quad {\bf 5}_1 \, , \quad {\bf 5}_{-1} \, , \quad {\bf 5}_0 \, , \quad {\bf 1}_1 \, , \quad {\bf 1}_2.
\end{align*}

\subsection{Models with more Mordell--Weil generators}\label{sec:more_MW_generators}

\subsubsection{Higher Mordell--Weil rank}

We have mentioned in the previous section that a higher rank $m$ of the Mordell--Weil group implies that there are possibly $m$ independent non-trivial central elements acting trivially on representations.
We will illustrate this now with concrete examples, in which the Mordell--Weil rank is 2. 
The simplest fibration that has two independent free sections arise from a generic cubic in an $\text{Bl}_2 \bbP^2 = dP_2$ fibration \cite{Borchmann:2013jwa,Cvetic:2013nia}.\footnote{A more general model with MW rank 2 has been recently constructed in \cite{Cvetic:2015ioa}; as shown there, the $\text{Bl}_2 \bbP^2$-fibration arises as a specialisation of this general construction.}
We will denote the divisor classes of the two sections $\sigma_{1,2}$ generating the Mordell--Weil group by $S_{1,2}$ and stick with the above notation of $Z$ being the zero section.

For simplicity, we focus on models with non-abelian gauge algebra $\mathfrak{su}(2)$, and label the single Cartan divisor by $E_1$.
All three such models arising from toric tops have been constructed in \cite{Lin:2014qga}.
Dubbed tops I, II and III, each of them turn out to have a different global gauge group structure, so it is instructive to analyse each individually.
In top I, the Shioda map $\varphi$ takes the sections to
\begin{align}
	\begin{split}
		& \varphi(\sigma_1) = S_1 - Z + \pi^{-1}(D_B) + \frac{1}{2} \, E_1 \, , \\
		& \varphi(\sigma_2) = S_2 - Z + \pi^{-1}(D_B') \, .
	\end{split}
\end{align}
Therefore, the $\fku(1)$ charges $(q_1, q_2)$ of $\mathfrak{su}(2)$ matter must satisfy $q_1 - \frac{1}{2} w \in \bbZ$ and $q_2 \in \bbZ$, where $w$ is the Dynkin label of $\mathfrak{su}(2)$ states.
Only the first condition leads to a central element acting non-trivially on $\mathfrak{su}(2)$ states. Clearly, it is an element of order 2, because $2\,q_1 - w \in \bbZ$.
We can also translate the second condition into a central element $C_2 = e^{2\pi i \, q_2} \in U(1)_2$. However, this element evidently just imposes charge quantisation $q_2 \in \bbZ$.
So the global gauge group structure is
\begin{align}\label{eq:global_gauge_group_su2I}
	G^\text{I} = \frac{SU(2) \times U(1)_1}{\bbZ_2} \times \frac{U(1)_2}{C_2} \cong \frac{SU(2) \times U(1)_1}{\bbZ_2} \times U(1)_2 \, .
\end{align}
The non-abelian part of the spectrum arranges consistently into
\begin{align*}
	{\bf 2}_{(\frac{1}{2}, -1)} \, , \quad {\bf 2}_{(\frac{1}{2}, 1)} \, , \quad {\bf 2}_{(\frac{1}{2}, 0)} \, .
\end{align*}

In the top II model, the Shioda map of the sections are
\begin{align}
	\begin{split}
		& \varphi(\sigma_1) = S_1 - Z + \pi^{-1}(D_B) + \frac{1}{2} \, E_1 \, , \\
		& \varphi(\sigma_2) = S_2 - Z + \pi^{-1}(D_B') + \frac{1}{2} \, E_1 \, .	
	\end{split}
\end{align}
Now both $\fku(1)$ charges must satisfy $q_i - \frac{1}{2}\, w \in \bbZ$. Put differently, there are now two central elements of order 2, 
\begin{align}\label{eq:central_elements_su2II}
\begin{split}
	C^\text{II}_1 = (\xi \times \mathds{1}) \otimes e^{2\pi i \, q_1} \otimes 1 \, , \\
	C^\text{II}_2 = (\xi \times \mathds{1}) \otimes 1 \otimes e^{2\pi i \, q_2} \, ,
\end{split}
\end{align}
of the covering group $\tilde{G} = SU(2) \times U(1)_1 \times U(1)_2$ that have to act trivially on all representations.
Each therefore generates a separate $\bbZ_2$ subgroup of $\tilde{G}$, leading to the actual gauge group
\begin{align}\label{eq:global_gauge_group_su2II}
	G^\text{II} = \frac{ SU(2) \times U(1)_1 \times U(1)_2}{ \bbZ^{(1)}_2 \times \bbZ^{(2)}_2} \, ,
\end{align}
where it needs to be understood that $\bbZ_2^{(i)}$ lies in the center of $SU(2) \times U(1)_i$.
Consequently, the $\mathfrak{su}(2)$ matter are charged as
\begin{align*}
	{\bf 2}_{(\frac{1}{2}, \frac{3}{2})} \, , \quad {\bf 2}_{(\frac{1}{2}, -\frac{1}{2})} \, , \quad {\bf 2}_{(\frac{1}{2}, \frac{1}{2})} \, .
\end{align*}

Finally, there is also the top III with Shioda map
\begin{align}
		\begin{split}
		& \varphi(\sigma_1) = S_1 - Z + \pi^{-1}(D_B)\, , \\
		& \varphi(\sigma_2) = S_2 - Z + \pi^{-1}(D_B') \, ,
	\end{split}
\end{align}
which clearly leads to trivial central elements. Hence, the gauge group in this case is just $G^\text{III} = SU(2) \times U(1)_1 \times U(1)_2$. 
The spectrum in this case is\footnote{
Note that we have included a completely uncharged doublet here that was previously missed in \cite{Lin:2014qga}. 
In fact, the codimension two locus of $I_3$ fibres corresponding to this matter was noticed. 
However, the monodromy around a codimension 3 sublocus interchanging two of the fibre components was misinterpreted as projecting out the matter states.
But due to the vanishing charges, this doublet is actually a real representation, i.e., the two fibre components are homologically equivalent. 
Thus the monodromy in higher codimension exchanging them is not surprising and actually expected geometrically.
A similar observation holds for singlets charged under a discrete $\bbZ_2$ symmetry \cite{Klevers:2014bqa, Mayrhofer:2014haa}.
}
\begin{align*}
	{\bf 2}_{(1,0)} \, , \quad {\bf 2}_{(1,1)} \, , \quad {\bf 2}_{(0,1)} \, , \quad {\bf 2}_{(0,0)}\, .
\end{align*}

Before we move on, let us briefly comment on a peculiar behaviour of the centre when we rotate the $\fku(1)$s. 
Concretely, it was noted in \cite{Lin:2014qga} that, if we redefine the $\fku(1)$ charges $(q_a, q_b) = (-q_1 , q_2 - q_1)$ in top II, the spectrum is identical to that of top I. In fact, this is a consequence of a toric symmetry relating tops I and II.
How is it compatible with the seemingly different gauge group structures \eqref{eq:global_gauge_group_su2I} and \eqref{eq:global_gauge_group_su2II}?
To understand this, let us rewrite the central elements \eqref{eq:central_elements_su2II} in terms of the rotated $\fku(1)$ charges.
Explicitly, we have $e^{2 \pi i \, q_1} \otimes 1 = e^{-2 \pi i \, q_a} \otimes 1$ and $1 \otimes e^{2 \pi i \, q_2} = 1 \otimes e^{2 \pi i \, (q_b - q_a)} = e^{-2\pi i \, q_a} \otimes e^{2 \pi i \, q_b}$.
So the central elements are
\begin{align}
\begin{split}
	& C^\text{II}_1 = (\xi \times \mathds{1}) \otimes e^{-2 \pi i \, q_a} \otimes 1 \, , \\
	& C^\text{II}_2 = (\xi \times \mathds{1}) \otimes e^{-2 \pi i \, q_a} \otimes e^{2 \pi i \, q_b} = C^\text{II}_1 \circ (\mathds{1} \otimes 1 \otimes  e^{2 \pi i \, q_b})\equiv  C_1^\text{II} \circ \tilde{C}^\text{II}_2\, ,
\end{split}
\end{align}
where here we use $\circ$ to denote the group multiplication in $SU(2) \times U(1)^2$.
Note that we are dealing with central elements, hence they all commute.
In the gauge group \eqref{eq:global_gauge_group_su2II} of top II, both $C^\text{II}_{1,2}$ must act trivially on all states.
The above equation implies that this is equivalent to $C^\text{II}_1$ and $\tilde{C}^\text{II}_2$ acting trivially.
But since $\tilde{C}^\text{II}_2$ lies in $U(1)_b$, we have 
\begin{align}\label{eq:global_gauge_group_su2II_rotated}
	G^\text{II} = \frac{SU(2) \times U(1)_a}{\langle C^\text{II}_1 \rangle} \times \frac{U(1)_b}{ \langle \tilde{C}^\text{II}_2 \rangle} \, .
\end{align}
Now the second quotient structure just imposes that the $U(1)_b$ charges are integer for all states, which is also implemented in \eqref{eq:global_gauge_group_su2I} by the second quotient.
Therefore, we have shown that by rotating the $\fku(1)$s in top II, the global gauge group structure \eqref{eq:global_gauge_group_su2II} (including charge quantisation) turns out to be equivalent to that of top I \eqref{eq:global_gauge_group_su2I}.

\subsubsection{Inclusion of Mordell--Weil torsion}

Let us now look at a model with Mordell--Weil group $\bbZ \oplus \bbZ_2$.
This example---studied extensively in \cite{Mayrhofer:2014opa} (and also appears in a slightly different fashion in \cite{Klevers:2014bqa})---has, in addition to the zero section a section $\sigma_f$ generating the free part and a section $\sigma_r$ generating the 2-torsional part of the Mordell--Weil group.
The fibration has two $\mathfrak{su}(2)$ factors with Cartan divisors $C$ and $D$, i.e., the covering gauge group is $SU(2)_C \times SU(2)_D \times U(1)$.
Under the Shioda map, the free section with divisor class $S$ maps onto
\begin{align}
	\varphi(\sigma_f) = S - Z + \pi^{-1}(D_B) + \frac{1}{2}\,C \, , 
\end{align}
giving rise to the central element $(\xi \times \mathds{1}) \otimes \mathds{1} \otimes e^{2\pi i \, q} \in \bbZ^{(f)}_2 \subset SU(2)_C \times U(1)$.
For the torsional section, one may determine the Shioda map analogously \cite{Mayrhofer:2014opa} through the conditions \eqref{eq:no_charge_generic_fibre} to \eqref{eq:no_charge_W_bosons}.
This yields
\begin{align}
	\varphi(\sigma_r) = V - Z + \pi^{-1}(D'_B) + \frac{1}{2}\, (C - D) \, .
\end{align}
Because of the 2-torsional property of $\sigma_r$ and $\varphi$ being a homomorphism, we know that $\varphi(\sigma_r) = 0$.
Analogous to the derivation of \eqref{eq:integer_pairing_condition} (with $q^{\bf w}=0$), it means that $\frac{1}{2}\,(w_C-w_D)$, with $w_{C,D}$ being the weights of $SU(2)_C \times SU(2)_D$ irreps, must be integral. 
So it defines another central element $\exp(\pi i (w_C - w_D))$ generating the `diagonal' $\bbZ_2^{(r)}$ of the $\bbZ_2 \times \bbZ_2$ centre of $SU(2)_C \times SU(2)_D$, which has to act trivially on representations of the F-theory compactification.
Therefore, the global gauge group structure is
\begin{align}
	\frac{SU(2)_C \times SU(2)_D \times U(1)}{\bbZ_2^{(f)} \times \bbZ_2^{(r)}} \, ,
\end{align}
where, in order for the notation to make sense, we need to clarify that $\bbZ_2^{(f)}$ acts only on $SU(2)_C \times U(1)$ representations, whereas $\bbZ_2^{(r)}$ acts on $SU(2)_C \times SU(2)_D$ representations.

Note that the $\bbZ_2^{(r)}$ quotient forbids any matter transforming as the fundamental representation under a single $SU(2)$ factor, irrespective of the $U(1)$ charge.
On the other hand, any matter transforming in the fundamental representation of $SU(2)_C$ must have $U(1)$ charge $\frac{1}{2} \mod \bbZ$ due to the $\bbZ_2^{(f)}$ quotient.
Thus, it is not surprising that the spectrum of the model contains, in addition to a charge 1 singlet, only bifundamental matter with charge $1/2$.

\subsection{F-theory Standard Models}\label{sec:examples_Standard_Models}

We now come to a class of somewhat more phenomenologically interesting models, namely elliptic fibrations realising the Standard Model gauge algebra $\fkg_\text{SM} = \mathfrak{su}(3) \oplus \mathfrak{su}(2) \oplus \mathfrak{u}(1)$ in F-theory.

The first model, presented in \cite{Klevers:2014bqa} (and labelled there as $X_{F_{11}}$) has $\fkg_\text{SM}$ as the full gauge algebra and the exact Standard Model spectrum (at the level of representations).
The inclusion of fluxes in \cite{Cvetic:2015txa} resulted in a first globally consistent three-chiral-family Standard Model construction in F-theory.
As mentioned in the introduction, the Standard Model spectrum is consistent with a global gauge group structure $(SU(3)\times SU(2) \times U(1))/\bbZ_6$.
With the new insights from section \ref{sec:general_story}, we can now explicitly show that in F-theory, we indeed can construct such a global structure.
In the $X_{F_{11}}$ model, the Shioda map of the free section is
\begin{align}\label{eq:shioda_map_XF11}
	\varphi(\sigma) = S - Z +\pi^{-1}(D_B) + \frac{1}{2} \, E_1^{\mathfrak{su}(2)} + \frac{1}{3} ( 2\,E_1^{\mathfrak{su}(3)} + E_2^{\mathfrak{su}(3)} ) \, ,
\end{align}
where $E_i^{\mathfrak{h}}$ denotes the Cartan generator(s) of the corresponding subalgebra $\mathfrak{h}$.\footnote{
Compared to \cite{Klevers:2014bqa}, we have switched the order of the $\mathfrak{su}(3)$ Dynkin labels by exchanging $E_1^{\mathfrak{su}(3)}$ and $E_2^{\mathfrak{su}(3)}$.
This exchanges the notion of ${\bf 3}$ and $\overline{\bf 3}$, making the charges identical to that of the Standard Model.
}
If we denote the weight vectors of $\mathfrak{su}(3)$ resp.~$\mathfrak{su}(2)$ by $(w_1, w_2)$ resp.~$\omega$ and the $\fku(1)$ charge by $q$, then the integrality condition \eqref{eq:integer_pairing_condition} for $X_{F_{11}}$ reads
\begin{align}\label{eq:integer_pairing_condition_XF11}
	q - \frac{1}{2}\omega - \frac{1}{3} (2\,w_1 + w_2) \in \bbZ \, .
\end{align}
Because $\omega, w_i \in \bbZ$, the smallest positive integer $\kappa$ such that $\kappa\,q \in \bbZ$ \textit{for all} possible charges $q$ is $\kappa = 6$.
Thus, the central element
\begin{align}
	C_{X_{F_{11}}} = \left[ e^{- 2\pi i \, \frac{2w_1 + w_2}{3} } \times \mathds{1}_{SU(3)} \right] \otimes \left[ e^{- 2\pi i \, \frac{\omega}{2} } \times \mathds{1}_{SU(2)} \right] \otimes e^{2\pi i \, q}
\end{align}
acting on $SU(3) \times SU(2) \times U(1)$ representations has order 6, so it defines a $\bbZ_6$ subgroup of the centre, i.e., the global gauge group is 
\begin{align}
	G_{F_{11}} = \frac{SU(3) \times SU(2) \times U(1)}{\bbZ_6} \, .
\end{align}
The condition \eqref{eq:integer_pairing_condition_XF11} implies that fundamental matter charged only under $\mathfrak{su}(3)$ must have charges $\frac{1}{3} \mod \bbZ$, while pure fundamentals of $\mathfrak{su}(2)$ have $q = \frac{1}{2} \mod \bbZ$. 
Inspecting the highest weight of the bifundamental, $\omega = 1, w_1 = 1, w_2 = 0$, we see that this representation must have charge $\frac{1}{6} \mod \bbZ$.
Correspondingly, the geometric spectrum,
\begin{align}
	({\bf 3,2})_{1/6} \, , \quad ({\bf 1, 2})_{-1/2} \, , \quad ({\bf 3,1})_{2/3} \, , \quad ({\bf 3, 1})_{-1/3} \, , \quad ({\bf 1,1})_1 \, ,
\end{align}
agrees with that of the Standard Model.

A different class of Standard-Model-like models was constructed in \cite{Lin:2014qga}, of which we have examined the $\mathfrak{su}(2)$ sector already above. The $\mathfrak{su}(3)$ sector is constructed in the analogous fashion with tops, which can be then combined with any $\mathfrak{su}(2)$ top to yield the non-abelian part of $\fkg_\text{SM}$.
Due to the rank 2 Mordell--Weil group, these models have an additional $\fku(1)$ symmetry, which can be used to implement certain selection rules.
As elaborated on in \cite{Lin:2014qga}, for each combination of the tops, there are again multiple ways of identifying the hypercharge $\fku(1)$ as a linear combination of the geometric $\fku(1)$s; the choice is tied to the role of the selection rule and the identification of the geometric spectrum with that of the Standard Model.
For definiteness, we focus on one particular choice of tops and identification, for which there also exists an extensive analysis including $G_4$-fluxes \cite{Lin:2016vus}.
In this case, the Shioda map, which for the first section also yields the hypercharge $\fku(1)$ generator, reads
\begin{align}
\begin{split}
	\fku(1)_{Y} : \quad & \varphi(\sigma_1) = S_1 - Z + \pi^{-1}(D_B) + \frac{1}{2} E^{\mathfrak{su}(2)}_1 + \frac{1}{3} (2E_1^{\mathfrak{su}(3)} + E^{\mathfrak{su}(3)}_2 ) \, , \\
	& \varphi(\sigma_2) = S_2 - Z + \pi^{-1}(D'_B) + \frac{1}{3} (2E_1^{\mathfrak{su}(3)} + E^{\mathfrak{su}(3)}_2 ) \, .
\end{split}
\end{align}
Analogously to the $X_{F_{11}}$ model, the first section leads to a central element of $SU(3) \times SU(2) \times U(1)_Y$ generating a $\bbZ^{(Y)}_6$ subgroup.
Meanwhile, the second section clearly generates a $\bbZ^{(2)}_3 \subset SU(3) \times U(1)_2$.
Hence, the global gauge group is
\begin{align}
	\frac{SU(3) \times SU(2) \times U(1)_Y \times U(1)_2}{\bbZ^{(Y)}_6 \times \bbZ^{(2)}_3} \, .
\end{align}
Consistently, the $( U(1)_Y , U(1)_2)$ charges of the fundamental representations arrange as follows:
\begin{itemize}
	\item $({\bf 3, 2})$ must have charge $(\frac{1}{6} \! \mod \bbZ, \frac{1}{3} \! \mod \bbZ)$.
	\item $({\bf 3,1})$ must have charge $(\frac{1}{3} \! \mod \bbZ, \frac{1}{3} \! \mod \bbZ)$.
	\item $({\bf 1,2})$ must have charge $(\frac{1}{2} \! \mod \bbZ, 0 \! \mod \bbZ)$.
\end{itemize}
Geometrically, the representations of the $\mathfrak{su}(3) \oplus \mathfrak{su}(2) \oplus \fku(1)_Y$ subalgebra agrees with that of the Standard Model (with additional singlets with no hypercharge).
However, the $\fku(1)_2$ charge discriminates between states that would be otherwise indistinguishable under the Standard Model algebra.
Note that, in order to get in touch with the actual Standard Model, one ultimately needs to lift the second $\fku(1)$ from the massless spectrum, which would require further investigation, see \cite{Lin:2016vus}.

\section{Relationship to (un-)higgsing}\label{sec:higgsing}

In this section, we want to explore the origin of the non-trivial global gauge group structure in higgsing processes.
In F-theory, $\fku(1)$s can often be unhiggsed into the Cartan of non-abelian gauge algebras \cite{Morrison:2012ei, Morrison:2014era, Klevers:2014bqa, Cvetic:2015ioa, Klevers:2016jsz}. 
Given that the Cartan charges of non-abelian matter are naturally integrally quantised, one might wonder if and how this is related to the restrictions on the $\fku(1)$ charges after breaking the non-abelian symmetry.

In fact, the F-theory Standard Model fibration $X_{F_{11}}$ we discussed in section \ref{sec:examples_Standard_Models} is shown in \cite{Klevers:2014bqa} to geometrically unhiggs into a Pati-Salam-like theory with $[SU(4) \times SU(2)^2]/\bbZ_2$ gauge group.
In this case, the $\bbZ_6$ centre of the Standard Model is known to arise from the representation theory of $[SU(4) \times SU(2)^2]/\bbZ_2$.\footnote{Note that the unhiggsed non-abelian group does not necessarily have to have a non-trivial global structure in order to induce one after breaking, cf.~$SU(5) \rightarrow [SU(3) \times SU(2) \times U(1)]/\bbZ_6$.}
One may ask whether it is possible to also unhiggs $X_{F_{11}}$ to an $SU(5)$ fibration.

However, the geometric description of unhiggsing is in general quite involved, since one does not a priori know the deformation corresponding to the specific unhiggsing process.
In order to gain some further intuition, we therefore restrict our analysis to a specific class of models, for which we have a good handle on the geometry.
For these, we show explicitly that the restrictions on the $\fku(1)$ charges leading to the global gauge group structure arise from a larger, purely non-abelian gauge theory.
As we will see, the unhiggsed non-abelian gauge algebra depends on the fibre split structure induced by the section.

\subsection[Unhiggsing the \texorpdfstring{$\fku(1)$}{u(1)} in a \texorpdfstring{$\text{Bl}_1 \bbP_{112}$}{Bl1P112} fibration]{Unhiggsing the \boldmath{$\fku(1)$} in a \boldmath{$\text{Bl}_1\bbP_{112}$} fibration}

The class of models we analyse have non-abelian gauge algebras engineered in a $\text{Bl}_1 \bbP_{112}$ fibration, a.k.a.~the Morrison--Park model \cite{Morrison:2012ei}.
Such models have a gauge algebra of the form $\fkg \oplus \fku(1)$.
Note that a broad class of such constructions has been classified through an analogue of Tate's algorithm in \cite{Kuntzler:2014ila}.
They can be realised as a toric hypersurface defined by the vanishing of the polynomial
\begin{align}\label{eq:general_MorrisonPark_hypersurface}
	P := w^2 \, s + b_0 \, w \, u^2 \, s+ b_1\,u\,v\,w\,s + b_2\,v^2\,w + c_0\,u^4 + c_1\,u^3\,v + c_2\,u^2\,v^2 + c_3\,u\,v^3 \, ,
\end{align}
where $[u:v:w]$ are the projective $\bbP_{112}$ coordinates.\footnote{
The authors of \cite{Morrison:2012ei} showed that, due to the constant coefficient in the $w^2\,s$-term, one can in fact absorb the terms with $b_0$ and $b_1$ through a coordinate redefinition, effectively setting them to 0. 
The inclusion of these terms allows for a more straightforward construction of non-abelian algebras, either via Tate's algorithm or via tops. 
Here, we have adopted the notation set in appendix B of \cite{Morrison:2012ei}, which is related to the notation of \cite{Kuntzler:2014ila} by exchanging $b_0 \leftrightarrow b_2$; also, the coordinates $(u,v,w)$ are labelled $(w,x,y)$ in \cite{Kuntzler:2014ila}.
}
Furthermore, $s$ is the blow-up coordinate whose vanishing defines the addition rational section generating the Mordell--Weil group.
As already discussed in \cite{Morrison:2012ei}, this fibration has a complex structure deformation, $b_2 \rightarrow 0$, which enhances the gauge algebra from $\fku(1)$ to an $\mathfrak{su}(2)_b$ localised over $c_3 = 0$.
In the absence of any additional non-abelian singularities, this enhancement can be understood as the inverse, i.e., unhiggsing, of breaking the $\mathfrak{su}(2)_b$ to $\fku(1)$ with its adjoint representation.
Under this breaking, the resulting singlets of the Morrison--Park model with charge 1 and 2, respectively, are remnants of fundamentals and adjoints, respectively, of the $\mathfrak{su}(2)_b$ theory.

By tuning the coefficients $b_i,c_j$ to vanish to certain powers along a divisor $\{\theta\}$ of the base, i.e., $b_i = b_{i,k} \, \theta^k$ and similarly for $c_j$, the fibres overs $\{\theta\}$ develop Kodaira singularities corresponding to a certain simple gauge algebra $\fkg$.
The above deformation still exists for these gauge enhanced models in the form of $b_{2,k} \rightarrow 0$ (we will abusively write $b_{2,k} \equiv b_2$).
This deformation will then modify the fibres over $c_{3,k'} \equiv c_3$ to have an $\mathfrak{su}(2)$ singularity, since a generic choice for $\{\theta\}$ will not affect this codimension one locus.
Likewise, the codimension two enhancement leading to the fundamentals of $\mathfrak{su}(2)$ will still persist in the presence of the divisor $\{\theta\}$, as it will generically not contain this codimension two locus.

\subsection[Charge constraints on \texorpdfstring{$\mathfrak{su}(2)$}{su(2)} matter from higgsing]{Charge constraints on \boldmath{$\mathfrak{su}(2)$} matter from higgsing}

\begin{figure}[ht]
\centering
	\def\svgwidth{.4\hsize}
	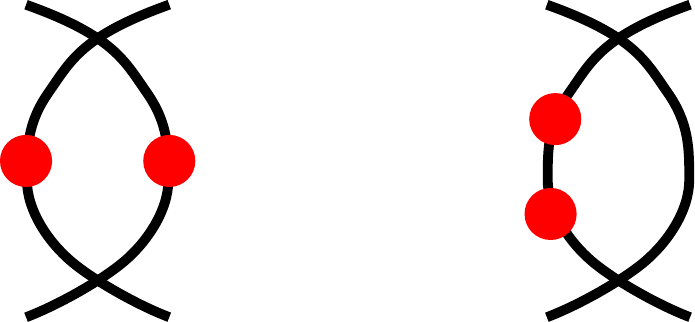
	\caption{The split types of an $I_2$-fibre by two sections (red dots): $(01)$ on the left and $(0|1)$ on the right.}
	\label{fig:fibre_split}
\end{figure}

To keep things simple, we will focus on the easiest example with $\fkg = \mathfrak{su}(2)_a$ from $I_2$-singularities over $\{\theta\}$, which we construct using `toric tops'. The subscript $a$ is to distinguish it from the $\mathfrak{su}(2)_b$ gauge algebra, which arises from unhiggsing the $\fku(1)$.
The spectrum of these models consists of singlets with charges 1 and 2, and fundamentals of $\mathfrak{su}(2)_a$.
The charges of these fundamentals depend on the fibre split, which for $\fkg = \mathfrak{su}(2)_a$ can only be of type $(01)$ or $(0|1)$ (see figure \ref{fig:fibre_split}).
With the techniques of \cite{Bouchard:2003bu}, one finds two different $\text{Bl}_1 \bbP_{112}$ fibrations with additional $\mathfrak{su}(2)_a$ singularities, corresponding to the $(01)$ and $(0|1)$ split type, respectively.

In the $(01)$ split model, whose resolved geometry is given by the vanishing of the polynomial
\begin{align}
	c_0\,e_0^2\,s^3\,u^4 + c_1\,e_0\,s^2\,u^3\,v + c_2\,s\,u^2\,v^2 + c_3\,e_1\,u\,v^3 + b_0\,e_0\,s^2\,u^2\,w + b_1\,s\,u\,v\,w + b_2\,e_1\,v^2\,w + s\,w^2 \, ,
\end{align}
the $\mathfrak{su}(2)$ fibre over $\{\theta\}$ is formed by the $\bbP^1$ fibres of the divisors $E_0 = \{e_0\}$ and $E_1 = \{e_1\}$.
Both the zero section $Z = [\{u\}]$ and the additional section $S = [\{s\}]$ intersect the component $\{e_0\}$.
Therefore the Shioda map of the section $S$ simply yields $S - Z$; correspondingly, we find $\mathfrak{su}(2)_a$ fundamentals with charges 1 and 0.
Their loci can be read of from the discriminant of this fibration:
\begin{align}\label{eq:discriminant_su2_01}
	\Delta^{(01)} \sim \underbrace{(b_1^2 - 4\,c_2)^2}_{\text{type $III$, no matter}} \, \underbrace{(b_2^2\,c_2 - b_1\,b_2\,c_3 + c_3^2)}_{{\bf 2}_1} \, \underbrace{[b_1^2\,c_0 - b_0\,b_1\,c_1 + c_1^2 + (b_0^2 - 4\,c_0)\,c_2]}_{{\bf 2}_0}\, \theta^2 + \CO( \theta^3) .
\end{align}
By tuning $b_2 \rightarrow 0$, we see from the discriminant \eqref{eq:discriminant_su2_01} that the locus of the ${\bf 2}_1$ curve now becomes $c_3^2$ --- precisely the locus of the $\mathfrak{su}(2)_b$ singularity:
\begin{align}\label{eq:discriminant_su2_01_unhiggsed}
	\tilde{\Delta}^{(01)} \sim \underbrace{(b_1^2 - 4\,c_2)^2}_{\text{type $III$, no matter}} \, \underbrace{[b_1^2\,c_0 - b_0\,b_1\,c_1 + c_1^2 + (b_0^2 - 4\,c_0)\,c_2]}_{P}\, \theta^2 \, c_3^2 + \CO( \theta^3 , c_3^3) \, .
\end{align}
Again, the first curve, $\{b_1^2 - 4\,c_2\}$, intersects both $\mathfrak{su}(2)_{a,b}$ divisors at codimension two loci of type $III$ enhancement, indicating the absence of any matter.
The intersections of the curve $\{P\}$ with $\{\theta\}$ and $\{c_3\}$ give rise to fundamentals of each $\mathfrak{su}(2)$ factor.
Finally, at the intersection $\{\theta\} \cap \{c_3\}$, we have bifundamentals of $\mathfrak{su}(2)_a \oplus \mathfrak{su}(2)_b$.
Since there are fundamentals of each $\mathfrak{su}(2)$ factor present, the global gauge group of the unhiggsed model must be $SU(2)_a \times SU(2)_b$.

Clearly, the higgsing in this case proceeds via adjoint breaking of $\mathfrak{su}(2)_b$, which preserves $\mathfrak{su}(2)_a$.
Geometrically, we immediately see that the uncharged ${\bf 2}_0$ matter in \eqref{eq:discriminant_su2_01} are completely unaffected by the \mbox{(un-)}higgsing process, as they arise from the $\mathfrak{su}(2)_a$ fundamentals along $\{\theta\} \cap \{P\}$ in \eqref{eq:discriminant_su2_01_unhiggsed}.
On the other hand, the charged fundamentals ${\bf 2}_1$ in \eqref{eq:discriminant_su2_01} arise from the bifundamentals sitting at $\{c_3\} \cap \{\theta\}$ before the higgsing; by the deformation that turns on $b_2$, they are localised at $\{b_2^2\,c_2 - b_1\,b_2\,c_3 + c_3^2\} \cap \{\theta\}$ after higgsing.
Hence, we can directly interpret the $\fku(1)$ after higgsing as the remnant $\mathfrak{su}(2)_b$ Cartan generator.
This explicitly identifies the $\fku(1)$ charge lattice with the weight lattice of $\mathfrak{su}(2)_b$.
In the `preferred normalisation', in which the singlets arising from higgsed remnants of $\mathfrak{su}(2)_b$ fundamentals and adjoints have charges 1 and 2, this implies that all $\fku(1)$ charges must be integer---which is equivalent to the statement of the integrality condition \eqref{eq:integer_pairing_condition} applied to the Shioda map $S-Z$.
In terms of the global gauge group structure, have $G_\text{glob} = SU(2) \times U(1)$.

The toric $(0|1)$ split $\mathfrak{su}(2)_a \oplus \fku(1)$ model is the vanishing of the polynomial
\begin{align}
	c_0\,e_0\,s^3\,u^4 + c_1\,e_0\,s^2\,u^3\,v + c_2\,e_0\,s\,u^2\,v^2 + c_3\,e_0\,u\,v^3 + b_0\,s^2\,u^2\,w + b_1\,s\,u\,v\,w + b_2\,v^2\,w + e_1\,s\,w^2 \, .
\end{align}
The $\mathfrak{su}(2)_a$ doublets have charges $3/2$ and $1/2$, consistent with the integrality condition \eqref{eq:integer_pairing_condition} for the Shioda map $S - Z + E_1/2$. Therefore, the global gauge group structure is $[SU(2) \times U(1)]/\bbZ_2$.
The discriminant of this fibration is
\begin{align}\label{eq:discriminant_su2_0|1}
	\Delta^{(0|1)} \sim \underbrace{(b_1^2 - 4 \,b_0\,b_2)^2}_{\text{type }III} \, \underbrace{b_2}_{{\bf 2}_{3/2}} \, \underbrace{Q}_{{\bf 2}_{1/2}} \, \theta^2 + \CO(\theta^3) \, ,
\end{align}
where $Q$ is a lengthy polynomial in the coefficients $b_i, c_j$.
However, since $b_2$ is now an explicit factor of the $\theta^2$ term of the discriminant, tuning $b_2 \rightarrow 0$ will clearly enhance the vanishing order of the discriminant in $\theta$.
Indeed, we find that after tuning, the discriminant becomes
\begin{align}\label{eq:discriminant_su2_0|1_unhiggsed}
\begin{split}
	\tilde{\Delta}^{(0|1)} & \, = \Delta^{(0|1)}|_{b_2=0} = c_3^2 \, \theta^3 \, \Delta_\text{res} \\
	& \, = - \underbrace{\theta^3}_{({\bf 3,2})} \,\underbrace{(b_1^2 - 4\,c_2\,\theta)^2}_{\text{type }III}\, \underbrace{(b_1^2\,c_0 - b_0\,b_1\,c_1 + b_0^2\,c_2 + c_1^2\,\theta - 4\,c_0\,c_2\,\theta)}_{P_1 \rightarrow ({\bf 1,2})}\,c_3^2 + \CO(c_3^3) \\
	& \, = -\underbrace{c_3^2}_{({\bf 3,2})}\, \underbrace{b_1^3}_{\text{type }IV}\, \underbrace{(b_1^3\,c_0 - b_0\,b_1^2\,c_1 + b_0^2\,b_1\,c_2 - b_0^3\,c_3)}_{P_2 \rightarrow ({\bf 3,1})}\, \theta^3 + \CO(\theta^4) \, ,
\end{split}
\end{align}
indicating an enhancement to $\mathfrak{su}(3)_a \oplus \mathfrak{su}(2)_b$.
At the intersection of the $\mathfrak{su}(3)_a$ divisor $\{\theta\}$ and the $\mathfrak{su}(2)_b$ divisor $\{c_3\}$ we naturally find bifundamentals ${\bf (3,2)}$.
Furthermore, the codimension two locus $\{P_1\} \cap \{c_3\}$ now supports fundamentals ${\bf (1,2)}$ of $\mathfrak{su}(2)_b$, and the locus $\{P_2\} \cap \{\theta\}$ supports fundamentals $\bf (3,1)$ of $\mathfrak{su}(3)_a$.

The deformation process of turning on the $b_2$ term now corresponds to bifundamental higgsing.
To see this, let us first look at the group theory.
The fundamental and adjoint representations decomposes as
\begin{align}
	\begin{split}
		\mathfrak{su}(3) \rightarrow \mathfrak{su}(2) \oplus \fku(1)_3 : \quad & {\bf 3} \rightarrow {\bf 2}_1 \oplus {\bf 1}_{-2} \, , \quad {\bf 8} \rightarrow {\bf 3}_0 \oplus {\bf 2}_3 \oplus {\bf 2}_{-3} \oplus {\bf 1}_0 \, ,\\
		\mathfrak{su}(2) \rightarrow \fku(1)_2 : \quad & {\bf 2} \rightarrow {\bf 1}_{-1} \oplus {\bf 1}_{1} \, , \quad {\bf 3} \rightarrow {\bf 1}_2 \oplus {\bf 1}_0 \oplus {\bf 1}_{-2} \, .
	\end{split}
\end{align}
Hence, the representations of the product algebra decompose according to
\begin{align}\label{eq:higgsing_spectrum_intermediate}
	\begin{split}
		\mathfrak{su}(3) \oplus \mathfrak{su}(2) \rightarrow \mathfrak{su}(2) \oplus \fku(1)_3 \oplus \fku(1)_2: \quad & ({\bf 3,2}) \rightarrow {\bf 2}_{(1,-1)} \oplus {\bf 2}_{(1,1)} \oplus {\bf 1}_{(-2,-1)} \oplus {\bf 1}_{(-2,1)} \, ,\\
		& ({\bf 3,1}) \rightarrow {\bf 2}_{(1,0)} \oplus {\bf 1}_{(-2,0)} \, ,\\
		& ({\bf 1,2}) \rightarrow {\bf 1}_{(0,-1)} \oplus {\bf 1}_{(0,1)} \, ,\\
		& ({\bf 8,1}) \rightarrow {\bf 3}_{(0,0)} \oplus {\bf 2}_{(3,0)} \oplus {\bf 2}_{(-3,0)} \oplus {\bf 1}_{(0,0)} \, ,\\
		& ({\bf 1,3}) \rightarrow {\bf 1}_{(0,2)} \oplus {\bf 1}_{(0,0)} \oplus {\bf 1}_{(0,-2)} \, .
	\end{split}
\end{align}
Therefore, by giving a vev to one of the singlets under the decomposition of the bifundamental, one breaks $\mathfrak{su}(3)_a \oplus \mathfrak{su}(2)_b$ to $\mathfrak{su}(2)_a \oplus \fku(1)$, where the $\fku(1)$ is a linear combination of the Cartans $\fku(1)_3$ and $\fku(1)_2$ such that the singlet receiving the vev is neutral under it.
This leaves the possibilities $\fku(1) = \fku(1)_2 \pm (\fku(1)_3/2)$, where we have chosen the normalisation such that the singlets after higgsing have charges 1 and 2.
It can be easily checked, that the two possibilities will in the end lead to the same $\mathfrak{su}(2)_a \oplus \fku(1)$ spectrum up to a sign for the $\fku(1)$ charge.
So by fixing $\fku(1) = \fku(1)_2 + (\fku(1)_3/2)$, we find
\begin{align}\label{eq:higgsing_spectrum_su3+su2_to_su2+u1}
	\begin{split}
		\mathfrak{su}(3) \oplus \mathfrak{su}(2) \rightarrow \mathfrak{su}(2) \oplus \fku(1) : \quad & ({\bf 3,2}) \rightarrow {\bf 2}_{-1/2} \oplus {\bf 2}_{3/2} \oplus {\bf 1}_{-2} \oplus {\bf 1}_0 \, ,\\
		& ({\bf 3,1}) \rightarrow {\bf 2}_{1/2} \oplus {\bf 1}_{-1} \, , \\
		& ({\bf 1,2}) \rightarrow {\bf 1}_{-1} \oplus {\bf 1}_{1} \, , \\
		& ({\bf 8,1}) \rightarrow {\bf 3}_0 \oplus {\bf 2}_{3/2} \oplus {\bf 2}_{-3/2} \oplus {\bf 1}_0 \, ,\\
		& ({\bf 1,3}) \rightarrow {\bf 1}_{2} \oplus {\bf 1}_0 \oplus {\bf 1}_{-2} \, .
	\end{split}
\end{align}
First, note that the charges agrees with the spectrum of the toric $(0|1)$ split $\mathfrak{su}(2)_a \oplus \fku(1)$.
Furthermore, comparing the matter loci \eqref{eq:discriminant_su2_0|1} to those of the unhiggsed $\mathfrak{su}(3) \oplus \mathfrak{su}(2)$ theory \eqref{eq:discriminant_su2_0|1_unhiggsed}, one can explicitly verify that the locus $\{Q\} \cap \{\theta\}$, supporting the ${\bf 2}_{1/2}$ matter of the $(0|1)$ model, decomposes upon unhiggsing into
\begin{align}
	\{Q\} \cap \{\theta\} \stackrel{b_2 \rightarrow 0}{\longrightarrow} \underbrace{\{c_3\} \cap \{\theta\}}_{(\bf 3,2)} \, \cup \, \underbrace{\{P_2\} \cap \{\theta\}}_{(\bf 3,1)} \, .
\end{align}
This confirms the geometric origin of the ${\bf 2}_{1/2}$ matter states we expected from the group theoretic higgsing process \eqref{eq:higgsing_spectrum_su3+su2_to_su2+u1}.
On the other hand, the codimension two locus of the ${\bf 2}_{3/2}$ matter, $\{b_2\} \cap \{\theta\}$, is promoted to the $\mathfrak{su}(3)$ divisor $\{\theta\}$, so, as expected, the adjoints of $\mathfrak{su}(3)$ contribute to ${\bf 2}_{3/2}$ upon higgsing.
The additional states originating from the bifundamentals are accounted for by explicitly checking the multiplicities.
Note that the unhiggsed theory also contains pure fundamentals of each gauge factor, hence the unhiggsed global gauge group must be $SU(3) \times SU(2)$.
So we conclude that the global gauge group structure, $[SU(2) \times U(1)]/\bbZ_2$, of the $(0|1)$ split model is a direct consequence of the bifundamental higgsing process of an $SU(3) \times SU(2)$ model, both field theoretically and geometrically in F-theory.

\subsection{Higher rank gauge algebras}

We have also repeated the above analysis for $(01)$ and $(0|1)$ split types with higher rank gauge algebras that appear in the classification of `canonical' Tate-like models in \cite{Kuntzler:2014ila}. This contains all A- and D-type algebras up to rank 5 as well as all exceptional algebras and $\mathfrak{so}(7)$.
We find that for any algebra $\fkg$ along $\{\theta\}$ with $(01)$ split, the tuning $b_2 \rightarrow 0$ never affects $\fkg$, i.e., the vanishing order of the discriminant along $\theta$ does not enhance further with this tuning.
It only leads to an $\mathfrak{su}(2)$ singularity along $\{c_3\}$, as we have seen before.
This is consistent with the tuning corresponding to adjoint (un-)higgsing of $\fku(1) \leftrightarrow \mathfrak{su}(2)$, as the Shioda map in these cases will always be $S-Z$, and hence the global gauge group being $G \times U(1)$.

For $(0|1)$ split type $\mathfrak{su}(n) \oplus \fku(1)$ models arising from $I_n$ singularities in codimension one, the tuning $b_2 \rightarrow 0$ unhiggs the model to $\mathfrak{su}(n+1) \oplus \mathfrak{su}(2)$.
A more general treatment of the group theoretic decomposition \eqref{eq:higgsing_spectrum_intermediate} and \eqref{eq:higgsing_spectrum_su3+su2_to_su2+u1} including 2-index anti-symmetric representations\footnote{
For $\mathfrak{su}(6) \oplus \mathfrak{su}(2) \rightarrow \mathfrak{su}(5) \oplus \fku(1)$, the inclusion of three-index anti-symmetric representations of $\mathfrak{su}(6)$ produces ${\bf 10}_{-3/5}$ states, in addition to the ${\bf 10}_{2/5}$ states that arises from two-index anti-symmetrics of $\mathfrak{su}(6)$. 
Those states, which also fit into the charge distribution of $(0|1)$ models (see \cite{Lawrie:2015hia}), arise in non-canonical models \cite{Mayrhofer:2012zy, Kuntzler:2014ila}.
}
confirms that the charges and global gauge group structure are consistent with bifundamental breaking.
For the other singularity types with $(0|1)$ split, we could verify that the tuning always enhances the singularity, i.e., increasing the rank of the gauge algebra along $\{\theta\}$ while still producing an $\mathfrak{su}(2)$ along $\{c_3\}$.
This suggests that the higgsing, that produces the non-trivial global gauge group structure for the $(0|1)$ type $\fkg \oplus \fku(1)$ model, is not achieved with adjoints.
However, to determine the exact matter content of the (un-)higgsed model requires a more detailed analysis, which we postpone to future works.

We also performed the same analysis for $(0||1)$ split types, i.e., where the section $S$ and zero-section $Z$ intersect next-to-neighbouring nodes of $\fkg$'s affine Dynkin diagram.
Naively, one finds an even higher enhancement along $\{\theta\}$ upon setting $b_2 \rightarrow 0$ (e.g., for $I_n$ singularities we find $\mathfrak{su}(n) \rightarrow \mathfrak{su}(n+2)$).
However, these tuned geometries always exhibit non-minimal codimension two loci (i.e., where the Weierstrass functions $(f,g,\Delta)$ vanish to orders $(4,6,12)$), indicating that there is---at least in F-theory compactifications to 6D--- hidden strongly coupled superconformal physics.
We hope to return to this issue in the future.

\section{A Criterion for the F-theory swampland}\label{sec:swampland}

We have seen that the geometry of elliptic fibrations imposes very stringent constraints on the $\fku(1)$ charges of matter states in F-theory compactifications.
A natural question that arises is if these constraints go beyond consistency conditions from a (supersymmetric) effective field theory (EFT) perspective.
Put differently, do they give rise to criteria for an EFT to be in the `swampland' \cite{Vafa:2005ui, ArkaniHamed:2006dz} of F-theory?
Given that in an EFT description, the global gauge group structure is often very obscure (e.g., because the spectrum of line operators is difficult to determine), it would be advantageous to have a criterion based solely on the gauge algebra $\fkg \oplus \bigoplus_k \fku(1)_k$ and the particle spectrum, which usually are directly accessible.

However, from the field theory perspective, there is no physically preferred normalisation for the $\fku(1)$ charge, whereas the integrality constraints appearing in F-theory models are only manifest in the geometrically preferred normalisation discussed in section \ref{sec:preferred_charge_normalisation}.
Thus, to formulate a swampland criterion based on charges, we first need to establish a method to fix the charge normalisation from the field theory perspective.
As we would like to argue now, singlet states (i.e., states uncharged under any non-abelian gauge symmetries) should provide such a reference for the normalisation.

\subsection{Singlet charges as measuring sticks}

Recall the geometrically preferred charge normalisation of F-theory discussed in section \ref{sec:preferred_charge_normalisation}.
This normalisation corresponds to having $\fku(1)$ generators $\omega_k = \varphi(\sigma_k)$ that arise from the normalised Shioda map $\varphi$ \eqref{eq:general_shioda_image} for free Mordell--Weil generators $\sigma_i$.
For simplicity, let us first look at the single $\fku(1)$ case.
In the preferred normalisation, singlets of $\fkg$ have integral charges, because their associated fibral curves satisfy $\Gamma \cdot E_i =0$, so $q^{\bf 1} = (S - Z) \cdot \Gamma \in \bbZ$.
In fact, we observe that in all $\fku(1)$ models with matter constructed so far in the literature, the charges of singlets computed with $\varphi(\sigma_k)$ for a free generator $\sigma_k$ of the Mordell--Weil group, are mutually relatively prime (there is at least one pair of coprime numbers).
This was expected to hold in \cite{Morrison:2012ei}, and viewed as a geometric incarnation of charge minimality \cite{Polchinski:2003bq, Banks:2010zn, Hellerman:2010fv, Seiberg:2011dr}, though a precise proof of this statement is to date not available.
If this statement holds in general, then singlets uniquely determine the preferred normalisation, since no rescaling of the $\fku(1)$ generator can preserve the charges to be integral and mutually relatively prime at the same time.

Observe that for a single $\fku(1)$, integer linear combinations of the singlet charges span $\bbZ$ if and only if the singlet charges are integer and mutually relatively prime.
This follows straightforwardly from elementary number theory, which says that $x,y$ are coprime integers if and only if there are integers $a,b$ such that $a\,x + b\,y = 1$.
The obvious generalisation to $m$ $\fku(1)$s is to require that a basis of $\fku(1)$ generators $\omega_k$ is in the preferred normalisation (i.e., arise as Shioda-maps $\omega_k = \varphi(\sigma_k)$ of free Mordell--Weil generators $\sigma_k$), if and only if the corresponding singlet charges are all integer and span the full integer lattice $\bbZ^m$.
Geometrically, this requirement is equivalent to say that the (Shioda-mapped) Mordell--Weil lattice is dual, with respect to the intersection pairing, to the lattice spanned by the fibral curves corresponding to singlets.
Similar to the case of a single $\fku(1)$, this condition is not proven in general.
For the purpose of this discussion, we will assume its validity, noting that it is true in all F-theory models with multiple $\fku(1)$s and charged matter constructed so far in the literature \cite{Borchmann:2013jwa, Borchmann:2013hta, Cvetic:2013nia, Cvetic:2013jta, Cvetic:2013uta, Cvetic:2013qsa, Cvetic:2015ioa, Krippendorf:2015kta}.

One may worry about cases where there are no singlets present at all, e.g., in F-theory models with non-higgsable $\fku(1)$s \cite{Martini:2014iza, Morrison:2016lix, Wang:2016urs}.
However, since we are interested in the interplay of non-abelian matter with the $\fku(1)$s, these particular models are not of concern because the tuning required for additional non-abelian algebra is expected to enhance the non-higgsable $\fku(1)$s into non-abelian symmetries.\footnote{We thank Wati Taylor and Yi-Nan Wang for pointing this out.}
Whether this phenomenon persists in all non-higgsable F-theory models, or if there are (higgsable) $\fku(1)$s with non-mutually relatively prime singlet charges, requires a more in-depth geometric analysis beyond the scope of this work.

Before we turn to the actual swampland conjecture, we would like to discuss how the normalisation condition carries over in cases where $\fku(1)$s are broken by either a higgsing or a fluxed-induced St\"uckelberg mechanism.
In the case of higgsing, i.e., giving a vacuum expectation value to a collection of massless states (in a D-flat manner), the breaking mechanism can be described geometrically by a complex structure deformation of the elliptic fibration.
This deformation yields another F-theory compactification, where the massless $\fku(1)$s are again realised by a non-trivial Mordell--Weil group.\footnote{
If there are remnant $\bbZ_k$ symmetries, then the complex structure deformation could yield a genus-one fibration without rational sections \cite{Braun:2014oya, Anderson:2014yva, Mayrhofer:2014haa, Mayrhofer:2014laa, Morrison:2014era}.
However, it is generally believed that there is always an elliptic fibration---the Jacobian fibration---with well-defined rational sections, giving the same F-theory.
}
Thus, the above formulation of the normalisation condition carry over directly.

On the other hand, a $G_4$-flux-induced breaking mechanism (which in F-theory is only possible in 4D and 2D) is not geometrised.
Hence, we have to understand field theoretically how the singlet charges behave in such a situation.
In the following, we will restrict our attention to four-dimensional compactifications, noting that the 2D case proceeds analogously \cite{Schafer-Nameki:2016cfr}.
In 4D, the field theoretic description of flux-induced breaking of $\fku(1)$s has been worked out in detail in type II (see \cite{MarchesanoBuznego:2003axu, Blumenhagen:2006ci, Plauschinn:2008yd} for a review) and subsequently in F-theory \cite{Cvetic:2012xn}.
In the latter setting, a non-zero flux induces a mass matrix
\begin{align}\label{eq:mass_matrix_flux_induced}
	M_{kl} = \sum_\alpha \xi_{k,\alpha} \, \xi_{l,\alpha} \quad \text{with} \quad \xi_{k,\alpha} = \int_{Y} G_4 \wedge \omega_k \wedge \pi^*J_\alpha \, ,
\end{align}
for the $\fku(1)$ gauge fields dual to the generators $\omega_k = \varphi(\sigma_k)$, $k=1,..., m$.
The flux-induced Fayet--Iliopoulos terms $\xi_{k,\alpha}$ are labelled by a basis $J_\alpha$ of $H^{1,1}(B)$, i.e., divisors on the base of the elliptic fourfold $\pi: Y \rightarrow B$.
Massless $\fku(1)$s are now precisely those linear combinations $\tilde{\omega}_s = \sum_k \lambda^s_{k} \, \omega_k$, which lie in the kernel of $M_{kl}$.
Due to \eqref{eq:mass_matrix_flux_induced}, this is equivalent to requiring 
\begin{align}\label{eq:masslessness_condition_flux}
	\forall \alpha \, : \quad \sum_i \xi_{k,\alpha} \, \lambda^s_k = 0 \, ,
\end{align}
which, depending on the $G_4$-flux, may or may not have non-trivial solutions.

Crucially, one can show that the FI-terms $\xi_{k,\alpha}$ can be taken to be integers due to the quantisation condition of $G_4$ (see appendix \ref{sec:app}).
Therefore, a non-trivial solution space $V$ of \eqref{eq:masslessness_condition_flux} can be generated by integer vectors $\lambda_k^s$, $s=1,..., \tilde{m} = \dim V$.
In other words, the massless $\fku(1)$s are generated by integer linear combinations $\tilde{\omega}_s = \lambda^s_k \, \omega_k = \varphi(\lambda^s_k \, \sigma_k)$ of the Mordell--Weil generators.
Hence, there must exist $\tilde{m}$ free Mordell--Weil generators $\tilde{\sigma}_s$ that span $V$.
Since the full singlet charge lattice was by assumption dual to the Mordell--Weil lattice, it must also contain the sublattice dual to $\Lambda = \text{span}_\bbZ (\tilde{\sigma}_s)$.
In other words, there is a basis $\tilde{\omega}_s = \varphi(\tilde{\sigma}_s)$ for the massless $\fku(1)$ generators, in which the singlet charges are all integer and their integer linear span fills out every lattice site of $\bbZ^{\tilde{m}}$.

So we have established that in an F-theory compactification with $m$ massless $\fku(1)$s, there are always $m$ free Mordell--Weil generators $\sigma_k$, such that the associated $\fku(1)$s are dual to the Shioda-divisors $\omega_k = \varphi(\sigma_k)$ given by \eqref{eq:general_shioda_image}.
In this geometrically preferred normalisation, the singlet charges are all integer and span the full $\bbZ^m$ lattice.
From the field theoretic point of view, which only has direct access to the singlet charges, it is crucial to realise that this normalisation is unique up to a unimodular transformation of the $\fku(1)$ generators, i.e., a change of basis for the Mordell--Weil (sub-)group.
To see that, let us denote by $q_k$, $k=1,...,m$ the charges of singlet states $Q_k$, which form a basis dual to $\omega_k$, i.e., $(q_k)_i = \delta_{ki}$.
Now suppose that we picked a different basis $\omega'_l$ for the $\fku(1)$ generators, in which the singlet charges again span $\bbZ^m$, with basis $q'_k$ dual to the $\omega'_k$.
While the $q_k$ and $q'_k$ correspond to different physical states $Q_k$ and $Q'_k$, their charge vectors both span $\bbZ^m$, so there must exist a change of basis, i.e., a unimodular matrix $U$, such that $Q'_k = U_{kl}\,Q_l$.
For the dual generators $\omega_k$ and $\omega'_k$, the corresponding transformation $\omega'_k = U^{-1}_{kl} \, \omega_l$ is then again unimodular.
Therefore, the sections $\sigma'_k = U^{-1}_{kl} \, \sigma_l$ generate the same lattice as $\sigma_k$, i.e., they are also Mordell--Weil generators.
Thus, the $\fku(1)$ generators $\omega'_k = \varphi(\sigma'_k)$ are also in the geometrically preferred normalisation.

\subsection{The swampland criterion}

Having established that the singlet charges provide a measuring stick for determining the preferred normalisation, we are now in a position to formulate the criterion which needs to be satisfied by EFTs arising from an F-theory compactification.
Given a theory with an unbroken $\fku(1)^{\oplus m} \oplus \fkg$ gauge symmetry, normalise the $\fku(1)$s such that the singlet charges are integer and span $\bbZ^m$. 
As argued above, we assume that this is always possible in F-theory.
In this case, the corresponding $\fku(1)$ generators $\omega_k$ are given by the Shioda-map \eqref{eq:general_shioda_image} of some free Mordell--Weil generators $\sigma_k$.
Then, due to \eqref{eq:L_of_rep} and \eqref{eq:integer_condition_charges_rep}, the difference $q_k^{(1)} - q_k^{(2)}$ of the $\fku(1)_k$ charges for any two representations ${\CR}^{(i)} = (q_k^{(i)} \, , {\CR}^{(i)}_\fkg)$ must be integer if $R^{(1)}_\fkg = R^{(2)}_\fkg$.
In other words, the condition states that singlets under the non-abelian gauge algebra provide a reference for the spacing of $\fku(1)$ charges, which has be respected also by all non-abelian matter in a given $\fkg$-representation, even if these may have fractional charges.
Any EFT that does not satisfy this criterion must lie in the F-theory `swampland', i.e., cannot be an F-theory compactification.

Note that this condition goes beyond anomaly cancellation.
An example in 6D is given by a tensorless $\mathfrak{su}(2) \oplus \fku(1)$ theory with 10 uncharged adjoint hypers, 64 fundamental hypers with charge 1/2, 8 fundamental hypers with charge 1, 24 singlet hypers with charge 1, and 79 uncharged hypers.
Clearly, the $\fku(1)$ is properly normalised according to our condition above, since there are only one type of charged singlets with charge 1.
However, despite the anomalies \cite{Park:2011wv} been cancelled with Green--Schwarz coefficients $a = -3, b_{\fku(1)} = 4, b_{\mathfrak{su}(2)} = 6$, the presence of both charge 1 and 1/2 $\mathfrak{su}(2)$ fundamentals does not meet our `swampland' criterion.
In 4D, the constraints are even weaker, since a completely vector-like spectrum is always gauge-anomaly-free, independent of charges or representations.

In summary, our necessary condition for an effective field theory with gauge algebra $\fku(1)^{\oplus m} \oplus \fkg$ to be an F-theory compactification requires to first establish a `preferred' normalisation.
This normalisation is determined by having all singlet charges being integer, and their integer span generates $\bbZ^m$.
Then, the difference of charges for matter in the same $\fkg$-representation must be integer.
Any field theory not satisfying this condition must lie in the F-theory swampland.
We re-emphasise that this criterion relies on the two key assumptions consistent with the current literature: (a) F-theory models with gauge algebra $\fku(1)^{\oplus m} \oplus \fkg$ and charged matter always have singlets, and (b) their charges in the preferred normalisation span the full integer lattice.
Were it not for these assumptions, we could always rescale the $\fku(1)$s so that all charges are integer, and the above conditions are trivially satisfied.
To sharpen our criterion will eventually require a rigorous proof of both assumptions.

Finally, we point out that our arguments are based on intersection properties of the Shioda map divisors with fibral curves, which in the F-theory compactification give rise to massless states in the effective field theory.
On the other hand, there are also massive states in string compactifications coming from higher Kaluza--Klein states or KK reductions along non-harmonic forms.
In light of the recent development of the weak gravity conjecture (see \cite{ArkaniHamed:2006dz, Cheung:2014vva, Montero:2015ofa, Heidenreich:2015nta, Heidenreich:2016aqi, Montero:2016tif, Palti:2017elp} for an incomplete list) which puts constraints on $\fku(1)$ charges of states relative to their masses, an interesting question is whether the global gauge group structure is also respected by massive states in F-theory.
If so, it would certainly be interesting to explore if and how our F-theory swampland arguments fit into more general quantum gravity concepts.

\section{Summary and outlook}\label{sec:summary}

In this work, we have shown that F-theory compactifications with an abelian gauge factor generically come equipped with a non-trivial gauge group structure $[G \times U(1)]/{\cal Z}$.
The finite subgroup ${\cal Z} = \bbZ_\kappa \subset {\cal Z}(G) \times U(1)$ of the centre is generated by an element $C$, which we have constructed explicitly from the Shioda map of the Mordell--Weil generator.
Geometrically, different centres ${\cal Z}$ arise from different fibre split types, i.e., relative configurations of the zero and the generating sections on the codimension one singular fibres determining $G$.
At the level of representations, the construction---generalising that for torsional sections \cite{Mayrhofer:2014opa}---imposes specific constraints on the allowed $\fku(1)$ charges $q_{\cal R}$  of each $G$-representation ${\cal R}$, such that ${\cal Z}$ acts trivially on $(q_{\cal R}, {\cal R})$.
These constraints can be equivalently viewed as a refined charge quantisation condition:
There is a normalisation \eqref{eq:general_shioda_image} of the $\fku(1)$ such that all charges of matter in a given $G$-representation $\cal R$ span a (one-dimensional) lattice with integer spacing.
A non-trivial gauge group structure is then reflected in a relative shift between charge lattices of different $G$-representations by multiples of $1/\kappa$.

We have exemplified our findings in several concrete models that have been constructed throughout the literature.
Using these examples, we have also demonstrated that the argument straightforwardly generalises to multiple $\fku(1)$ factors, i.e., higher rank Mordell--Weil groups, and also to cases with both free and torsional sections.
Each generator (free or torsional) leads to an independent central element (possibly trivial), such that in general, ${\cal Z}$ is a product of $\bbZ_{\kappa_k}$ factors.
In particular, when applied to the `F-theory Standard Models' \cite{Lin:2014qga, Klevers:2014bqa, Cvetic:2015txa, Lin:2016vus}, we found that these models realise the physical Standard Model gauge group $[SU(3) \times SU(2) \times U(1)_Y]/\bbZ_6$.
Correspondingly, the geometric spectrum completely agrees with the physical representations of the $\mathfrak{su}(3) \oplus \mathfrak{su}(2) \oplus \fku(1)_Y$ algebra.

We have also explored the connections of $\fku(1)$ charge restrictions to the process of unhiggsing into a larger non-abelian group.
Relying on simple class of geometries with $\mathfrak{su}(2) \oplus \fku(1)$ gauge algebra, we have shown explicitly that the two different global gauge group structures unhiggs, both geometrically and field theoretically, into different non-abelian gauge groups.
Concretely, geometries with gauge group $SU(2) \times U(1)$ arise as adjoint higgsing of $SU(2) \times SU(2)$, whereas $[SU(2) \times U(1)]/\bbZ_2$ is a result of bifundamental higgsing of $SU(3) \times SU(2)$.
Note that the non-abelian gauge groups in which we unhiggs do not have any non-trivial structure, i.e., there are no torsional sections.
In general, models in the same class with gauge group $[G \times U(1)]/{\cal Z}$ unhiggs under the same complex structure deformation into $G' \times SU(2)$ with $G \subseteq G'$; equality holds only if ${\cal Z} = \{1\}$.

However, for $[G \times U(1)]/{\cal Z}$ models, where the zero section and the Mordell--Weil generator do not intersect the same or neighbouring fibre components in codimension one, the unhiggsing procedure introduces codimension two non-minimal loci.
In compactifications on a threefold, one would usually associate such a non-minimal locus with the existence of tensionless strings and interpret it as a superconformal sector of the 6D field theory.
Clearly, it would be exciting to investigate how superconformal physics enters the global gauge group structure, and also gain insight into 4D compactifications, where non-minimal loci are less understood.
It is worth pointing out that the centre also plays a crucial role in the higgsing of $\fku(1)$s to discrete symmetries \cite{Garcia-Etxebarria:2014qua, Klevers:2014bqa}.
With the explicit description of the centre laid out in these notes, we look forward to apply our new insights to phenomenologically more appealing models with discrete symmetries \cite{SM_Z2}.

We have also studied how the non-trivial global gauge group structures give rise to an F-theory `swampland' criterion.
Formulated in terms of $\fku(1)$ charges, a field theory with gauge algebra $\fkg \oplus \fku(1)$ that arises from F-theory must satisfy the following condition.
By normalising the $\fku(1)$ such that all singlet charges are integer and span the full integer lattice, charges of matter having the same non-abelian $\fkg$-representation, which individually can be fractional, must differ from each other by integers.
Our analysis also shows that this criterion generalises to the case of multiple $\fku(1)$s, which remain massless in the low energy effective theory after a higgsing and/or turning on $G_4$-flux.
While this condition is stronger than cancellation of field theory anomalies and hence can be used to rule out swampland theories, their validity is based on the assumption that in F-theory compactifications, the singlets serve as a `measuring stick' for the $\fku(1)$ charges.
Geometrically, it is based on the observation that any F-theory model with $\fku(1)$s and charged matter have in particular charged singlets, whose corresponding fibral curve in the elliptic fibration spans a lattice that is dual to the (free) Mordell--Weil lattice under the intersection pairing.
Note that this observation extends the conjecture \cite{Morrison:2012ei} made for a singlet $\fku(1)$, that singlet charges computed with respect to the normalised Shioda-map \eqref{eq:general_shioda_image} are integer and mutually relatively prime.
To make the swampland criterion precise will therefore require a careful analysis of the intersection structures between sections and codimension two fibres in elliptically fibred Calabi--Yau manifolds.
Nevertheless, it would be interesting to study the connection of this swampland criterion to other quantum gravity conditions such as the weak gravity conjecture and extensions thereof.

\subsection*{Acknowledgement}

We are indebted to Timo Weigand, Craig Lawrie, Thomas Grimm, Wati Taylor, Yi-Nan Wang, and Cumrun Vafa for valuable discussions and comments on the draft.
Furthermore, we would also like to thank Paul Oehlmann, Eran Palti, Riccardo Penco and Fabian Ruehle for useful communications.
This work was supported in part by DOE Award DE-SC0013528 (M.C.~, L.L.), and by the Fay R.~and Eugene L.~Langberg Endowed Chair (M.C.) and the Slovenian Research Agency (M.C.).

\appendix

\section{Integrality of fluxed-induced Fayet--Iliopoulos terms}\label{sec:app}

In this appendix, we show that the fluxed-induced FI-terms \eqref{eq:mass_matrix_flux_induced} can be taken to be integers in \eqref{eq:masslessness_condition_flux}, which determines the massless linear combinations of $\fku(1)$s.
First, recall the consistency conditions \cite{Dasgupta:1999ss} for $G_4$-flux in F-theory compactified on an elliptic fourfold $\pi: Y \rightarrow B$:
\begin{align}\label{eq:G4_transversality_conditions}
\begin{split}
	& G_4 \in H^{2,2}(Y) \quad \text{such that} \quad  \\
	& \forall D_B \in H^{1,1}(B) \, \, \text{and} \, \, \forall C_B \in H^{2,2}(B): \, \int_Y G_4 \wedge Z \wedge \pi^* D_B = \int_Y G_4 \wedge \pi^* C_B = 0 \, ,
\end{split}
\end{align}
where $Z$ is the cohomology form dual to the divisor of the zero-section.
These so-called transversality conditions ensure that the flux preserves 4D Lorentz symmetry in the M-/F-theory duality.
Furthermore, in order not to break the non-abelian gauge algebra $\fkg$ with Cartan divisors $E_i$, we impose 
\begin{align}\label{eq:G4_gauge_symmetry_condition}
	\forall D_B \in H^{1,1}(B) : \quad \int_Y G_4 \wedge E_i \wedge \pi^* D_B \, .
\end{align}
Because of these conditions, the fluxed-induced FI-terms \eqref{eq:mass_matrix_flux_induced} for a $\fku(1)$ generator $\omega_k = \varphi(\sigma_k)$, with $\sigma_k$ a Mordell--Weil generator with divisor class $S_k$, simplifies to
\begin{align}\label{eq:simplified_FI-terms}
	\xi_{k,\alpha} = \int_Y G_4 \wedge \omega_k \wedge \pi^* J_\alpha = \int_Y G_4 \wedge (S_k - Z + l_i\,E_i + \pi^* D_B) \wedge \pi^*J_\alpha = \int_Y G_4 \wedge S_k \wedge \pi^* J_\alpha \, .
\end{align}

The basis elements $J_\alpha$ for the K\"ahler form $J_B$ of the base can be taken to be integer cohomology forms dual to divisor.
Then we can invoke the quantisation condition \cite{Witten:1996md},
\begin{align}
	G_4 + \frac{1}{2} c_2(Y) \in H^4(Y, \bbZ) \, ,
\end{align}
where $c_2(Y)$ is the second Chern class of $Y$.
Therefore, since the section classes $S_k$ are also integer, the FI-terms \eqref{eq:simplified_FI-terms} satisfy
\begin{align}\label{eq:integer_condition_FI-terms}
	\bbZ \ni \int_Y \left( G_4 + \frac{1}{2} c_2(Y) \right) \wedge S_k \wedge \pi^* J_\alpha = \xi_{i,\alpha} + \frac{1}{2} \int_Y c_2(Y) \wedge S_k \wedge \pi^* J_\alpha \, .
\end{align}
Because $c_2(Y) \in H^4(Y, \bbZ)$, the last term is at most half-integer.
In other words, $2\xi_{k,\alpha} \in \bbZ$ for all $k$ and $\alpha$.
Since an overall factor of the FI-terms does not affect the solutions of the masslessness condition \eqref{eq:masslessness_condition_flux} for $\fku(1)$s, we can thus assume from the beginning that $\xi_{k,\alpha} \in \bbZ$.


\bibliography{FTheory}{}
\bibliographystyle{JHEP} 

\end{document}